\documentclass[reprint,amsmath,amssymb, aps, prb,
floatfix,nofootinbib]{revtex4-1} 
\usepackage{graphicx}
\usepackage{dcolumn}
\usepackage{bm}
\usepackage{caption}
\usepackage{amssymb,amsmath,color}
\usepackage{url}
\usepackage{float}
\usepackage{braket}
\usepackage{hyperref} 
\usepackage{comment}
\hypersetup{colorlinks = true, linkcolor = blue, anchorcolor = blue, citecolor = blue, filecolor = blue, urlcolor = blue}
\usepackage{xcolor}
\usepackage{empheq}
\usepackage{enumitem}

\setlist[itemize]{
    leftmargin=*,
    itemsep=0pt,
    topsep=2pt,
    parsep=0pt
}

\begin{document}
\title{Moiré trapping of quadrupolar excitons in van der Waals trilayers}
\author{
  Indrajit Maity\textsuperscript{1,*},  
  Arash A. Mostofi\textsuperscript{2},
  \'{A}ngel Rubio\textsuperscript{3,4},
  Johannes Lischner\textsuperscript{2} \\ [1ex]
  \textsuperscript{1}Chemistry, School of Natural and Environmental Sciences, Newcastle University, Newcastle upon Tyne, NE1 7RU, UK.\\  
  \textsuperscript{2} Departments of Materials and Physics and the Thomas Young Centre for Theory and Simulation of Materials, Imperial College London, South Kensington Campus,
London, SW7 2AZ, UK. \\
  \textsuperscript{3} Max Planck Institute for the Structure and Dynamics of Matter, Luruper Chaussee 149, 22761 Hamburg, Germany. \\
  \textsuperscript{4} Initiative for Computational Catalysis and Center for Computational Quantum Physics, Flatiron Institute, Simons Foundation, New York City, New York 10010, USA.\\ 
  \textsuperscript{*}Corresponding author: indrajit.maity02x@gmail.com
}
\date{\today}

\begin{abstract}
\section*{Abstract}
Quadrupolar excitons in van der Waals heterostructures - quantum superpositions
of anti-aligned dipolar excitons - offer a platform to explore exotic
many-body physics, with applications to quantum sensing and photonic devices.
Yet their internal structure, symmetry, and real-space localisation remain
largely unknown. Here, we reveal the atomic-scale structure of quadrupolar
excitons in twisted WSe$_{2}$/WS$_{2}$/WSe$_{2}$ trilayers by solving the
Bethe-Salpeter equation within a large-scale atomistic framework. We discover
that large atomic relaxations at small twist angles give rise to two distinct
quadrupolar excitons trapped at moir\'e lattice sites, differing in the
in-plane symmetry of the electron density about the hole: one azimuthally
symmetric, with the density maximal at the hole, and one threefold symmetric,
with a node at the hole. Moir\'e trapping, neglected in commonly used models of
quadrupolar exciton formation, is critical to their many-exciton phases.
Without moir\'e trapping, quadrupolar excitons transition into anti-parallel
dipolar excitons on a bipartite square lattice, while with trapping, the same
dipoles are confined to a triangular lattice and experience geometric
frustration. Our study uncovers the highly non-trivial nature of quadrupolar
excitons, with direct implications for simulating frustrated quantum magnetism
in a fully tunable excitonic platform.
\end{abstract}

\maketitle

\section*{Introduction}
Multipole expansions encode the symmetry of a source through its moments, with
applications spanning molecular electrostatics~\cite{Buckinghampermanent1967},
gravitational radiation~\cite{Thornemultipole1980}, and many-body
physics~\cite{Suzukifirst2018,Santinimultipolar2009}. For example, in
electrostatics, dipoles arise from separated positive and negative charges,
while quadrupoles appear in more symmetric arrangements where both the total
charge and dipole moment vanish. Recently, this framework has found a striking
realization in van der Waals
heterostructures~\cite{Yuobservation2023,Lianquadrupolar2023,Liquadrupolar2023,Baievidence2023,Xiebright2023,Chenbichromatic2025,Jasinskiquadrupolar2025,Xietransition2024,Fengquadrupolar2025,Zhangdiverse2024,Sungbroken2020,Zhenyucontinuously2026,Gimsymmetric2025,Zhuobservation2025}.
By stacking and twisting two-dimensional (2D) materials, experimentalists have
demonstrated tunable moments of excitons, which are bound electron–hole pairs.
For example, excitons in symmetric trilayer transition metal dichalcogenides
such as
WSe$_{2}$/WS$_{2}$/WSe$_{2}$~\cite{Yuobservation2023,Lianquadrupolar2023} and
WS$_{2}$/WSe$_{2}$/WS$_{2}$~\cite{Liquadrupolar2023} show a transition from
quadrupolar to dipolar character when an out-of-plane electric field is
applied. The absence of a dipole moment leads to qualitatively distinct
interaction physics, making these systems an attractive platform for exploring
exotic many-body
phases~\cite{Xucorrelated2025,Guotopological2025,Mengstrong2025} and
optoelectronic applications~\cite{Liutunable2025,Zhenglocalization2023}. To
date, the dipolar and quadrupolar characters have been inferred almost
\textit{exclusively} from the electric-field dependence of photoluminescence
(PL), which shifts from linear to non-linear
behavior~\cite{Barrequadrupolar2023}. Direct real-space imaging of their
structure, symmetry, and twist-angle dependence remains elusive.

Accurate theory is essential to close this gap, provide atomistic insight, and
guide future experiments. Yet, theoretical modeling remains scarce. This
scarcity is mainly due to the prohibitive computational cost of
state-of-the-art many-body perturbation theory methods, such as the
GW-plus-Bethe-Salpeter equation (GW-BSE) approach, for simulating multilayer
twisted 2D materials with hundreds to thousands of
atoms~\cite{Blasebethe2020,Rohlfingelectron2000,Onidaelectronic2002,Louiediscovering2021}.
As a result, GW–BSE calculations have been limited to lattice-matched trilayer
systems containing only a few atoms~\cite{Deilmannquadrupolar2024,
Slobodkinquantum2020, Zhangquadrupolar2025, Qinquadrupolar2025,
Fengquadrupolar2025}. However, stacking and twisting drive pronounced atomic
rearrangements that \textit{qualitatively} alter exciton behavior. Such effects
are entirely absent in the lattice-matched models. For example, in
$\mathrm{WS_2/WSe_2}$ bilayers, atomic rearrangements give rise to localized
\textit{intralayer} excitons in $\mathrm{WSe_2}$ with strong electron-hole
overlap and charge-transfer character~\cite{Naikintralayer2022,
Sandhyahyperspectral2022, Maityorigin2026, Dalalsignatures2026,
Jinobservation2019, Wangtwist2025}. Structural relaxations are therefore
essential for any predictive description of multipolar excitons and their
field-tunable moments. Furthermore, accurate parameters characterizing
individual quadrupolar and dipolar excitons are essential input for modeling
the many-exciton phase diagram~\cite{Slobodkinquantum2020,
Astrakharchikquantum2021}. Their importance is evident from a striking
contradiction between theory and experiment~\cite{Yuobservation2023}: in the
exciton density regime where experiments observe quadrupolar excitons, models
that neglect structural relaxations instead predict a staggered dipolar phase.

Here, we overcome these challenges and provide the first moir\'{e}-scale
atomistic description of the electric-field-driven quadrupolar-to-dipolar
transition in twisted trilayer WSe$_{2}$/WS$_{2}$/WSe$_{2}$, fully accounting
for structural relaxations. To this end, we develop a general transfer-matrix
formalism for electron–hole Coulomb interactions in twisted multilayer 2D
materials and solve the fully atomistic BSE using ab initio electronic
structure as input to determine the exciton spectrum and its real-space
structure. We demonstrate that quadrupolar excitons form in
WSe$_2$/WS$_2$/WSe$_2$ trilayers with parallel outer WSe$_2$ layers, regardless
of the twist angle with WS$_{2}$. At small twist angles, large atomic
rearrangements trap quadrupolar excitons at moiré sites, splitting them into
two species that differ in the in-plane symmetry of the electron density about
the hole: one azimuthally symmetric with maximal density at the hole, and one
threefold symmetric with a node at the hole. Moir\'{e} trapping offers a
mechanism that reconciles previous theory with
experiment~\cite{Yuobservation2023}. By confining excitons to a triangular
lattice, it induces geometric frustration that stabilises the quadrupolar phase
across the experimentally observed density range.

\section*{Results}

\noindent \textbf{Bethe-Salpeter equation in multilayers.} We construct a fully
atomistic BSE Hamiltonian using Wannier functions obtained from large-scale
density functional theory (DFT) calculations of arbitrarily twisted
multilayers~\cite{Maityatomistic2025} \begin{widetext} \begin{equation}
\begin{split} \langle cv{\bf k}| \mathcal{H}_{\mathrm{BSE}}
|c^{\prime}v^{\prime}{\bf k^{\prime}}\rangle = (\epsilon_{c{\bf k}} -
\epsilon_{v{\bf k}}) \delta_{cc^\prime} \delta_{vv^\prime} \delta_{\bf
kk^\prime} -\frac{1}{N} \sum_{{\bf R},n_{1},n_{2}} {\big(C^{\bf
k}_{n_{1}c}\big)}^* C^{\bf k^\prime}_{n_{1}c^\prime} {C^{\bf k}_{n_{2}v}}
{\big(C^{\bf k^\prime}_{n_{2}v^\prime}\big)}^* W({\bf R} + ({\bf t}_{n_2} -
{\bf t}_{n_1})) e^{i ({\bf k - k^\prime})\cdot{\bf R}} \\ +\frac{1}{N}
\sum_{{\bf R},n_{1},n_{2}} {\big(C^{\bf k}_{n_{1}c}\big)}^* {C^{\bf
k}_{n_{1}v}} C^{\bf k^\prime}_{n_{2}c^\prime} {\big(C^{\bf
k^\prime}_{n_{2}v^\prime}\big)}^* V({\bf R} + ({\bf t}_{n_{2}} - {\bf
t}_{n_{1}})).  \label{bsewannier}
\end{split}
\end{equation}
\end{widetext}

Here, $\epsilon_{c(v)\mathbf{k}}$ are the quasiparticle energies of conduction
(valence) states, $C_{nm}^{\mathbf{k}}$ are the Wannier function coefficients,
$\mathbf{t}_n$ is the position of the $n$-th Wannier function in the home unit
cell, $\mathbf{R}$ are lattice vectors, $N$ is the number of
$\mathbf{k}$-points sampled in the Brillouin zone, and $W$ and $V$ are the
screened and bare Coulomb interactions. Details of the derivation of
Eqn.~\ref{bsewannier} are detailed elsewhere~\cite{Maityatomistic2025}. The key
innovation is that $W$ is obtained from the multilayer electrostatic problem
via a transfer-matrix formalism. This contrasts with previous approaches to
excitons in multilayers that also use classical electrostatics for screening
but were limited to dipolar interlayer excitons in lattice-matched bilayers and
treated within effective-mass frameworks~\cite{Cavalcanteelectrostatics2018,
Danovichlocalized2018, Oveseninterlayer2019,
Brunettioptical2018,Vaninterlayer2018}. This goes beyond our previous
work~\cite{Maityatomistic2025}, where the screening was restricted to the
analytical Rytova-Keldysh form and could not be generalised to arbitrary
twisted multilayer systems.

\begin{figure*}
    \centering
    \includegraphics[width=1.0\linewidth]{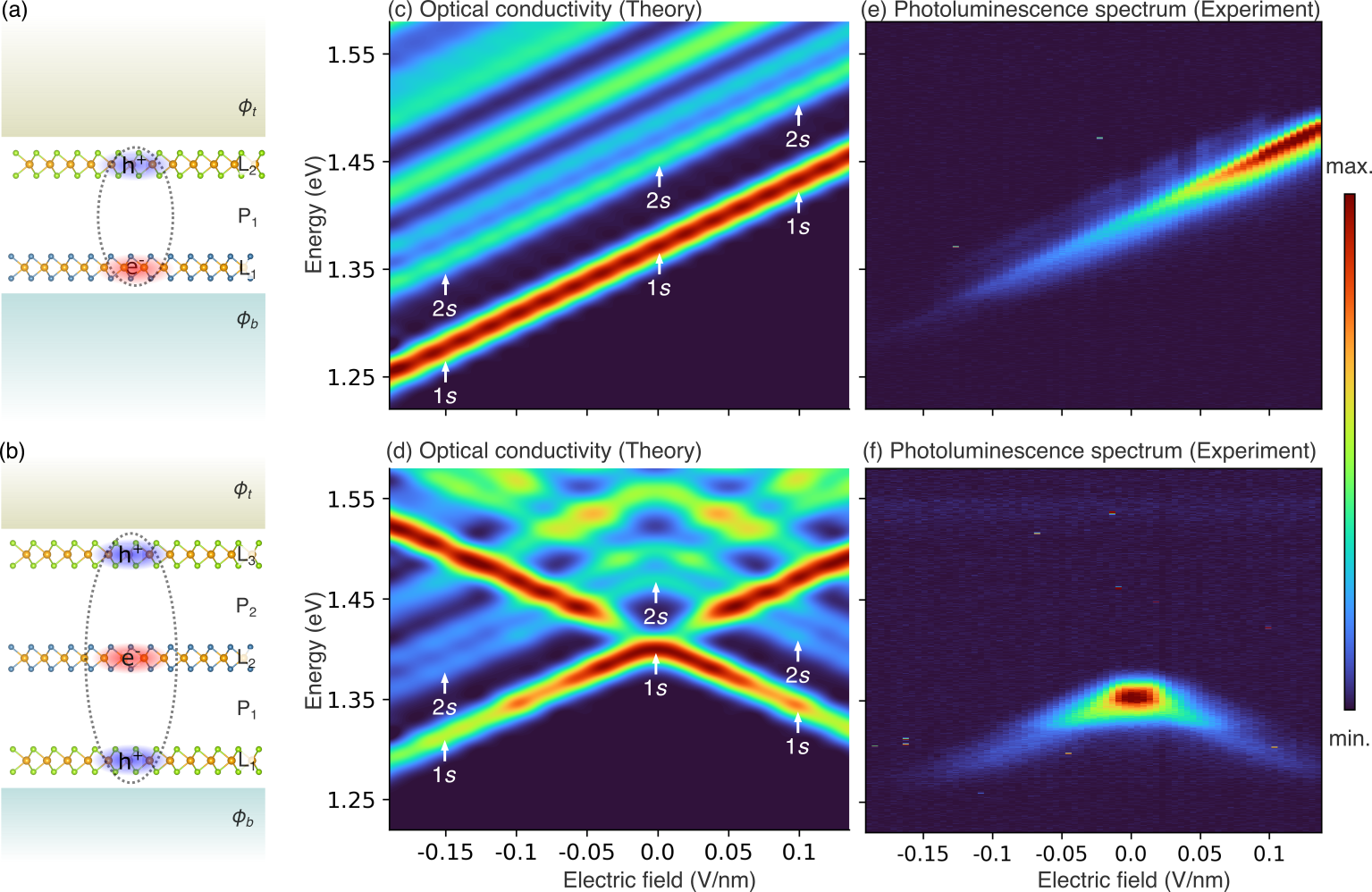}
    \caption{\textbf{Structure and out-of-plane electric-field response of
dipolar and quadrupolar excitons.} (a), (b): Schematics of dipolar and
quadrupolar excitons in WSe$_2$/WS$_2$ and WSe$_2$/WS$_2$/WSe$_2$
heterostructures sandwiched between substrates. The top and bottom
electrostatic potentials, $\phi_t$ and $\phi_b$, are related by the layer
($L_i$), propagation ($P_i$), and electron source matrices. (c),(e): Dipolar
excitons exhibit a linear Stark effect, whereas (d),(f) quadrupolar excitons
transition from a quadratic to a linear Stark response as the electric field
strength increases. The colorbar indicates both optical conductivity and
photoluminescence. Optical conductivities are from lattice-matched structures.
Experimental data adapted with permission from Yu \textit{et
al}~\cite{Yuobservation2023} (Copyright 2023 Springer Nature).} \label{fig1}
\end{figure*}

To illustrate the transfer-matrix method, we consider two 2D layers with
arbitrary stacking (Fig.~\ref{fig1}(a)). We first construct the screened
Coulomb potential in momentum space $\phi_{ij}(q)$ in layer $j$ due to an
electron in layer $i$. Along the out-of-plane direction $z$, we express the
potential in each region as a sum of exponentials $e^{\pm qz}$ and enforce
boundary conditions at each interface by combining propagation ($P_{m}$), layer
($L_{m}$), and source ($S$) matrices. Here, $P_m(d)$ propagates the potential
across the spacing $d$ between the $m$-th and $(m+1)$-th layers, $L_m(\chi_m,
\epsilon_1, \epsilon_2)$ describes the response of the $m$-th layer with
polarizability $\chi_m$ between dielectrics $\epsilon_1$ and $\epsilon_2$, and
$S$ is the electron source (see Fig.~\ref{fig1}(a)). The transfer matrix
determines the top potential from the bottom one, $\phi_{i=1,t}(q) = L_2 P_1
L_1 \left[\phi_{i=1,b} + S\right](q)$. From this we propagate through the
layers to find $\phi_{i=1,j}(q)$.
We repeat this for the second layer, determining $\phi_{i=2,t}(q) = \left[L_2
P_1 L_1\phi_{i=2,b} + S\right](q)$ and propagating to find $\phi_{i=2,j}(q)$.
The same approach extends naturally to the three-layer system (Fig.~\ref{fig1}(b)),
\begin{subequations}\label{all_sources}
\begin{align}
\phi_{i=1,t}(q) &= L_{3}P_{2}L_{2}P_{1}\Big[L_{1}\phi_{i=1,b} + S\Big](q), \\
\phi_{i=2,t}(q) &= L_{3}P_{2}\Big[L_{2}P_{1}L_{1}\phi_{i=2,b} + S\Big](q), \\
\phi_{i=3,t}(q) &= \Big[L_{3}P_{2}L_{2}P_{1}L_{1}\phi_{i=3,b} + S\Big](q).
\end{align}
\end{subequations}
See Sec.~A of the Supplementary Information (SI) for details and generalization
to arbitrary multilayers. We use $\chi_{\mathrm{WS_2}}=75.73~\mathrm{\AA}$ and
$\chi_{\mathrm{WSe_2}}=90.18~\mathrm{\AA}$ obtained from
DFT~\cite{Berkelbachtheory2013}. Finally, we apply a Hankel transform to map
$\phi_{ij}(q)$ to the real-space $W$ entering the BSE Hamiltonian,
Eq.~\ref{bsewannier}.

\noindent \textbf{Quadrupolar excitons in lattice-matched trilayers.} WS$_{2}$
and WSe$_{2}$ have a lattice mismatch of \(4\%\), producing a moir\'{e} pattern
with about 8 nm lattice constant even in the absence of
twist~\cite{Yuobservation2023,Jinobservation2019,Sandhyahyperspectral2022,Mengstrong2025}.
To disentangle the moir\'e-related effects, we first construct lattice-matched
WSe$_2$/WS$_2$ bilayers and WSe$_2$/WS$_2$/WSe$_2$ trilayers 
with a lattice constant of 0.325~nm. In both systems, WSe$_2$ and WS$_2$ are 
antiparallel ($\theta = 60^\circ$). In the trilayer, we keep the two outer 
WSe$_2$ layers parallel (AA stacking, $\theta = 0^\circ$). Both systems have type-II band 
alignment, with the valence band maximum in WSe$_2$ and the conduction band 
minimum in WS$_2$, and a direct band gap at the $K$ point.
Fig.~\ref{fig1}(c),(d) compare the calculated optical conductivity of the two
systems under an applied out-of-plane electric field, $E_{z}$. To place all
results on a common energy reference, we apply a rigid scissor shift that
aligns the lowest-energy bright exciton to 1.4 eV at zero field. The same shift
is used throughout. The bilayer $1s$ interlayer exciton shows a linear Stark
effect over the entire field range, consistent with a permanent dipole. In
contrast, the trilayer $1s$ interlayer exciton shows a quadratic response up to
$E_{z}=\pm 50$ mV/nm, crossing over to a linear regime at higher fields. This
quadratic-to-linear crossover is the hallmark of a field-induced transition
from quadrupolar to dipolar excitons~\cite{Barrequadrupolar2023}.
Fig.~\ref{fig1}(c),(d) also show higher-energy exciton Rydberg states up to the
$2s$ level. At zero field, the $2s$–$1s$ energy splitting decreases from 82 meV
in the bilayer to 68 meV in the trilayer, reflecting enhanced screening in the
trilayer. We also observe Rydberg excitons beyond the $2s$ state, but the
calculations used here are not sufficiently converged to accurately determine
their energies and oscillator strengths (See Sec.~B of the SI for details on
interlayer and intralayer excitons). Fig.~\ref{fig1}(e),(f) show measured PL
spectra for structures where WSe$_{2}$ and WS$_{2}$ form a 8 nm moir\'{e}
pattern and the outer WSe$_{2}$ layers are nearly
aligned~\cite{Yuobservation2023}. Our calculations reproduce the key features
of the experimentally observed lowest bright excitons in PL spectra. We now
turn to moiré systems arising from the $\approx 4\%$ lattice mismatch between
WS$_2$ ($a = 0.319$\,nm) and WSe$_2$ ($a = 0.332$\,nm).

\begin{figure*}[htb!] 
    \centering
    \includegraphics[width=1.0\linewidth]{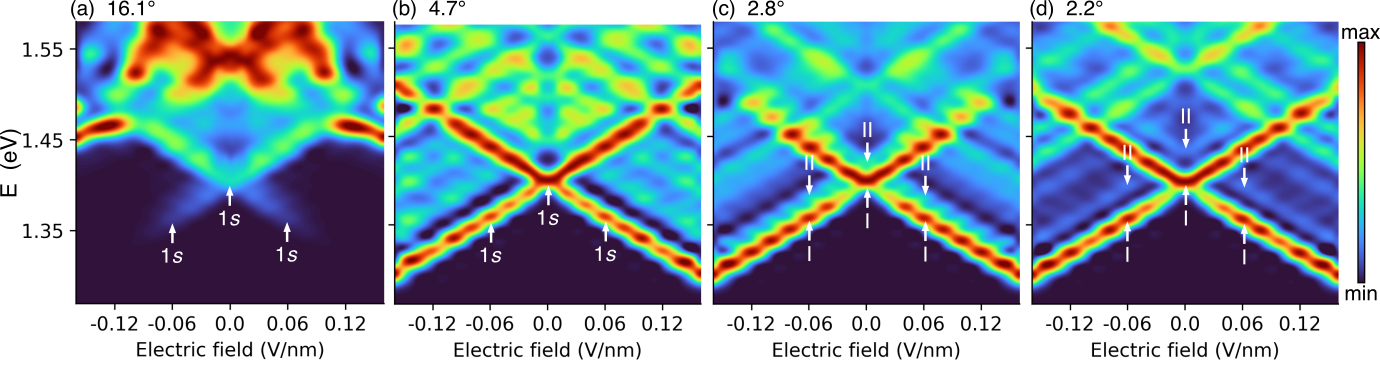}
    \caption{\textbf{Out-of-plane electric-field response of excitons in
twisted WSe$_{2}$/WS$_{2}$/WSe$_{2}$.} (a–d) As the electric field increases
from zero, the lowest bright exciton changes from a quadratic to a linear
response, indicating a transition from quadrupolar to dipolar behaviour. Peak I
and II in (c) and (d) are due to moiré trapping of quadrupolar excitons. The
colorbar indicates the optical conductivities.} \label{fig2}
\end{figure*}

\noindent \textbf{Twist-angle dependence of quadrupolar excitons.}
Fig.~\ref{fig2} shows the twist-angle dependence of the lowest-energy excitons
in WSe$_2$/WS$_2$/WSe$_2$ under different out-of-plane electric fields. The two
WSe$_2$ layers remain aligned, while the middle WS$_2$ layer is twisted by an
angle $\theta$ relative to the bottom WSe$_2$ layer. Regardless of twist angle,
the lowest bright exciton exhibits a quadratic-to-linear Stark response as
$|E_z|$ increases from zero. A zoom-in of the lowest excitons near zero field
highlighting the quadratic region is also shown in Fig.~\ref{fig4}(b). At large
twist angles (see Fig.~\ref{fig2}(a)), the peak intensities are weak, but they
increase as the twist angle decreases. At small twist angles (see
Fig.~\ref{fig2}(c),(d)), we find two quadrupolar excitons, labeled I and II,
with exciton II lying within 50~meV of exciton I. These peaks are
\textit{fundamentally} different from the $1s$ state in the lattice-matched
system without a moiré pattern. They correspond to moiré-trapped quadrupolar
excitons and are the main result of this work.

\begin{figure*}[ht!]
    \centering
    \includegraphics[width=1.0\linewidth]{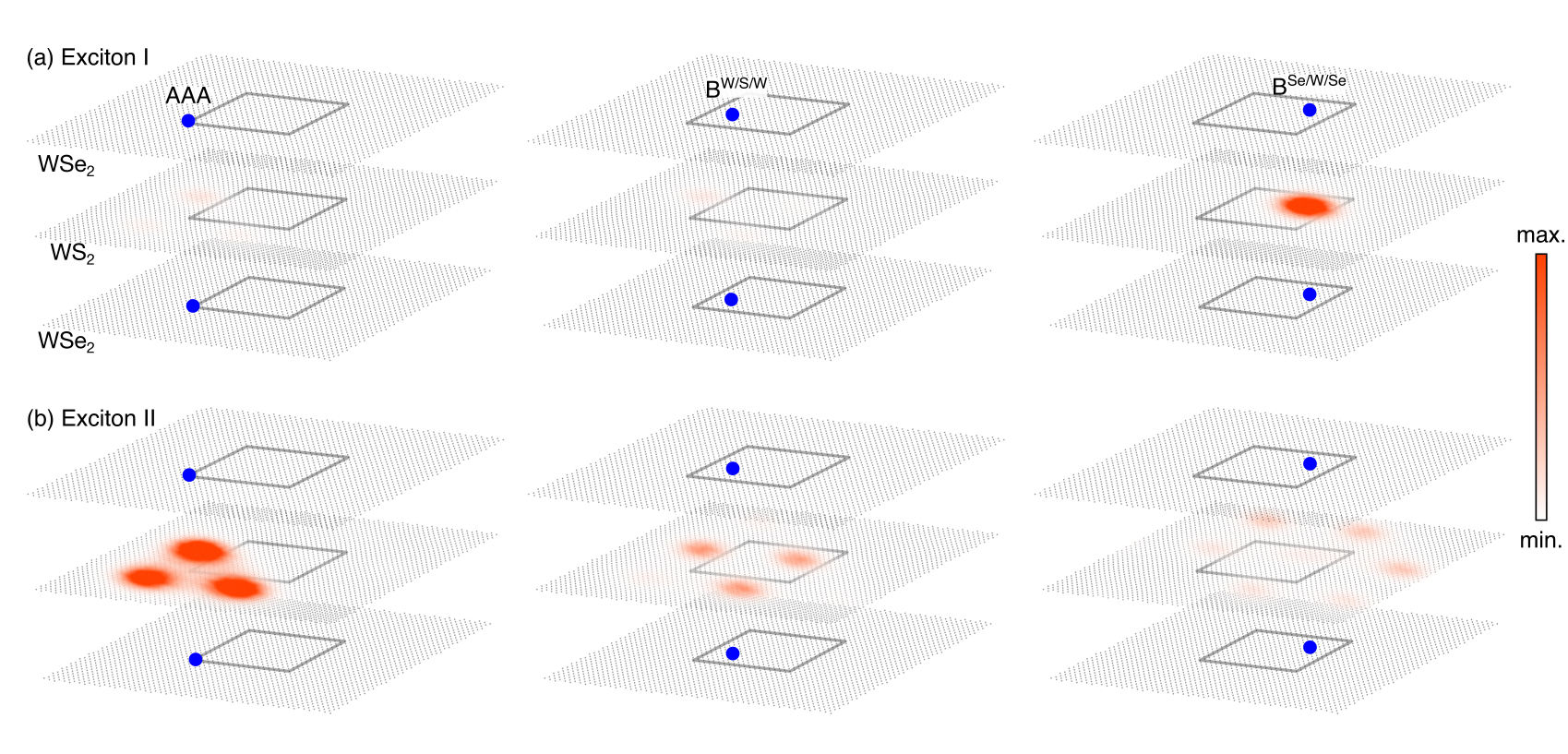}
    \caption{\textbf{Moir\'{e} quadrupolar exciton wavefunctions in a
2.2$^\circ$ twisted WSe$_2$/WS$_2$/WSe$_2$.} (a) Exciton I shows azimuthally
symmetric hole density. (b) Exciton II shows threefold symmetric electron
density, laterally displaced from the hole location. Red dots indicate the
electron positions, while solid lines show the moir\'e unit cell. The colorbar
indicates the hole densities. When the electron density in the WSe$_2$ layers
is fixed, the hole density is negligible and therefore not shown.} \label{fig3}
\end{figure*}

\begin{figure}
\centering
\includegraphics[width=1.0\linewidth]{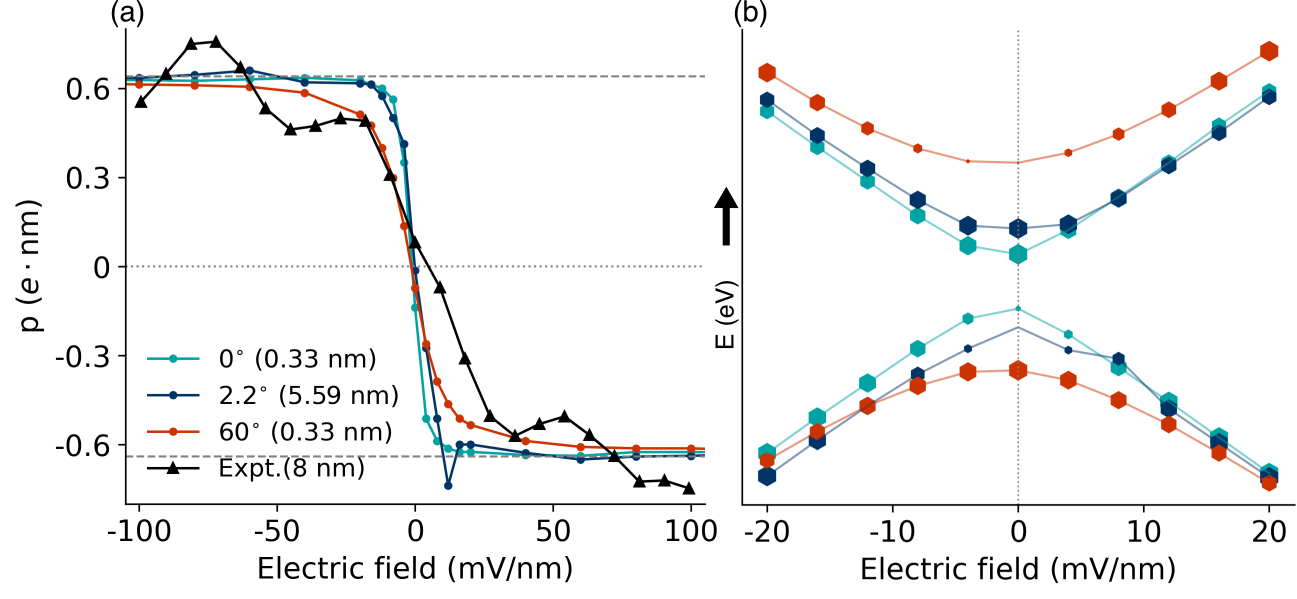}
\caption{\textbf{Tunable dipole moments and avoided crossings in twisted
WSe$_{2}$/WS$_{2}$/WSe$_{2}$.} Field-dependent (a) dipole moments of the lower
branch and (b) avoided-crossing gaps between the upper and lower branches for
several twist angles. The center of the gap is aligned across all systems.
Hexagon size indicates the oscillator strength of the excitons in (b).
Experimental data adapted with permission from Yu \textit{et
al}~\cite{Yuobservation2023} (Copyright 2023 Springer Nature).} \label{fig4}
\end{figure}

To demonstrate moiré trapping, we analyze the exciton wavefunctions associated
with excitons I and II at zero electric field. Fig.~\ref{fig3} shows the
electron distributions at a 2.2$^\circ$ twist with the hole fixed at different
stacking regions (See SI, Sec.~K for the corresponding hole wavefunctions at
fixed electron positions). For both excitons, the electron is located in WS$_2$
and the holes are in the WSe$_2$ layers. As shown in Fig.~\ref{fig3}(a),
exciton I has the electron and hole vertically aligned with the electron
density showing circular symmetry. The electron density is strongest in the
$\mathrm{B^{Se/W/Se}}$ stacking region, corresponding to Bernal stacking with
Se atoms of both WSe$_2$ layers directly above W atoms of WS$_2$. In contrast,
Fig.~\ref{fig3}(b) shows exciton II has the electron distributed across three
lobes at the vertices of a triangle, displaced laterally from the hole
position. The electron density is strongest in the AAA stacking configuration,
where the W and Se atoms in WSe$_2$ are directly on top of the W and S atoms in
WS$_2$, respectively. Crucially, exciton II has reduced rotational symmetry,
$C_3$, generating an in-plane octupolar moment absent in exciton I (see SI,
Sec.~C for further analysis). The exciton trapping in different stacking
regions, absent in lattice-matched trilayers and at large twist angles, arises
from moiré-induced atomic rearrangements. In the energetically unfavorable AAA
stacking region, atoms move anticlockwise in WSe$_2$ and clockwise in WS$_2$ to
reach a more stable configuration (see SI, Sec.~D for more details). The
symmetric WSe$_2$ layers on either side of WS$_2$ amplify these rearrangements
in WS$_2$. This produces a much stronger confining potential for electrons than
in a twisted bilayer system, consistent with previous
experiments~\cite{Zhenglocalization2023}. As a result, quadrupolar excitons
become trapped at approximately below $3^\circ$ twist, corresponding to moiré
lengths $\gtrsim 5$~nm. See Sec.~F of the SI for a comparison of the
quadrupolar excitons in the unrelaxed and relaxed atomic structures of the
2.2$^\circ$ twisted WSe$_{2}$/WS$_{2}$/WSe$_{2}$, which reveals that atomic
relaxation qualitatively changes the exciton localization.

A complementary approach to understanding exciton trapping due to atomic
rearrangements is to compare the in-plane strains that develop during atomic
relaxation. At a small twist angle ($2.2^\circ$), large in-plane strains of up
to $\sim\pm1\%$ develop in each layer, leading to trapped quadrupolar excitons.
In contrast, a large twist angle ($16.1^\circ$) shows negligible strains
($\lesssim 0.05\%$), and the excitons remain untrapped. See SI, Sec.~F for the
strain distributions at all twist angles.

\noindent \textbf{Two-dipole avoided crossing model.} To explain the electric
field-driven quadrupole-to-dipole transition, we use an \textit{avoided
crossing} model. In this model, two degenerate dipolar excitons with energy
$\mathcal{E}_d$ and dipole moments $\pm p_0$ are coupled via a quantum
tunneling strength $t$ under an applied electric field $E_z$. The system is
described by the Hamiltonian~\cite{Deilmannquadrupolar2024, Yuobservation2023,
Liquadrupolar2023, Xiebright2023, Lianquadrupolar2023} \begin{equation}
\begin{pmatrix}
\mathcal{E}_{d} + p_0 E_z & -t \\
-t & \mathcal{E}_d - p_0 E_z
\end{pmatrix}.
\label{avoided_H}
\end{equation}
Eqn.~\ref{avoided_H} has two eigenvalues, $\mathcal{E}_{q}^{\pm}= \mathcal{E}_d
\pm \sqrt{p_0^2 E_z^2 + t^2}$, which correspond to two quadrupolar exciton
branches. Since experiments have primarily observed the lower-energy
branch~\cite{Liquadrupolar2023}, we extract $p_0$ and $t$ by fitting the
field-dependent dipole moment $p(E_{z})$ from our BSE
calculations. This branch corresponds to the highlighted $1s$ state at large
twist angles and lattice-matched systems, and to peak I at small twist angles
(Fig.~\ref{fig2}). We fit $p(E_z)$ using
\begin{equation}
p(E_z) = -\frac{d\mathcal{E}_{q}^{-}}{dE_z}
= \frac{p_0^{2} E_z}{\sqrt{(p_0 E_z)^2 + t^2}}.
\label{sigmoid}
\end{equation}
At strong fields ($p_0 E_z \gg t$), the dipole saturates to $\pm p_0$, while at
weak fields ($p_0 E_z \ll t$), it vanishes, recovering the quadrupolar
character. Fig.~\ref{fig4}(a) compares our calculated $p(E_z)$ for all moiré
structures and the lattice-matched trilayer with
experiment~\cite{Yuobservation2023}, showing qualitative agreement (See SI
Sec.~L for results at additional twist angles). Table~\ref{table1} lists the
extracted parameters (See SI, Sec. I for example fit). The tunneling strength
$|t|$ increases monotonically as the twist angle decreases from $16.1^\circ$ to
$2.2^\circ$. The splitting at zero field $(\mathcal{E}_{q}^{+} -
\mathcal{E}_{q}^{-})$ in Table~\ref{table1}, obtained from the BSE
calculations, is consistent with the model prediction of $2|t|$.

Another key distinguishing factor between quadrupolar and dipolar excitons is
the field-dependent PL radiative rate,
$\gamma^{\text{rad}}$~\cite{Liquadrupolar2023, Yuobservation2023}. We capture
this behavior using Fermi’s golden rule~\cite{Palummoexciton2015} together with
the computed oscillator strengths,
$\gamma_{\text{dip}}^{\text{rad}}/\gamma_{\text{quad}}^{\text{rad}} \approx
\mathcal{P}_{E_z=20\,\text{meV/nm}}/\mathcal{P}_{E_z=0\,\text{meV/nm}}$.
Fig.~\ref{fig4}(b) shows the oscillator strengths of both exciton branches,
defined as $\mathcal{P} = \left|\sum_{cv{\bf k}} A^{S}_{cv{\bf k}}
    \langle v{\bf k} | p_x | c{\bf k} \rangle \right|^2$.
At $60^\circ$, $\mathcal{P}$ is highest for the lower branch, while at
$0^\circ$ it is highest for the upper branch. The bright quadrupolar exciton
near zero $E_z$ has a radiative rate $\gamma^{\text{rad}}$ 
1.5--1.8 times higher than the dipolar branch at high $E_z$, while the dark 
branch is 4--8$\times10^{3}$ times lower, depending on twist angle. Our results
agree well with the experimentally reported 1.7-fold faster radiative decay of
the bright quadrupolar exciton branch~\cite{Liquadrupolar2023}.

\begin{table}[h!]
\centering
\begin{tabular}{c c c c}
\hline
Lattice constant (nm) & $p_0$ (e.nm) & $|t|$ (meV) & \big[$\mathcal{E}_{q}^{+} - \mathcal{E}_{q}^{-}$\big] (meV) \\
\hline\hline 
0.325 (60$^\circ$) & 0.61 $\pm 0.02$ & 7.4 $\pm 1.6$ & 14.5 \\
0.325 (0$^\circ$) & 0.63 $\pm$ 0.02 & 2.6 $\pm 0.8$ & 3.8 \\ \\

1.15 (16.1$^\circ$) & 0.59  $\pm 0.02$ & 1.3 $\pm$ 0.7 & 1.8 \\
3.49 (4.7$^\circ$) & 0.66 $\pm 0.02$ & 1.4 $\pm$ 0.8 &  2.0\\
4.89 (2.8$^\circ$) & 0.64 $\pm$ 0.02 & 3.4 $\pm$ 0.9 & 6.5 \\
5.59 (2.2$^\circ$) & 0.64 $\pm$ 0.02 & 3.7 $\pm$ 0.9 & 6.9 \\
\hline
\end{tabular}
\caption{\textbf{Parameters of the avoided crossing model.} We obtain
$\mathcal{E}^{\pm}_{q}$ from our Bethe-Salpeter equation calculations and
extract $p_{0}$ and $t$ by fitting $\frac{d\mathcal{E}_{q}^{-}}{dE_{z}}$ to the
avoided crossing model. The twist angle is defined between neighboring WSe$_2$
and WS$_2$ layers.}
\label{table1}
\end{table}

\noindent \textbf{Many-exciton phases at zero $T$.} To demonstrate the impact
of moiré trapping, we map the zero-temperature phase diagram of interacting
quadrupolar excitons, both free and trapped. We model excitons as point-like
bosons and neglect their kinetic energy~\cite{Slobodkinquantum2020}. Building
on the avoided crossing model, the phases are governed by three competing
energy scales: electrostatic interactions, the energy cost
$(\mathcal{E}_{q}^{+}-\mathcal{E}_{q}^{-})/2$ of occupying the dipolar instead
of the quadrupolar state, and the moiré trapping energy $V_\text{M}$. Each
dipolar exciton carries an Ising degree of freedom $\sigma_i = \pm 1$ (up or
down), which selects the branch of the pair potential,
\begin{equation}
  V_{ij}(r) = f
  \begin{cases}
    \dfrac{2}{r} - \dfrac{2}{\sqrt{r^2 + d^2}},
    & \sigma_i\sigma_j = +1, \\[8pt]
    \dfrac{1}{r} + \dfrac{1}{\sqrt{r^2 + 4d^2}}
      - \dfrac{2}{\sqrt{r^2 + d^2}},
    & \sigma_i\sigma_j = -1,
  \end{cases}
\end{equation}
, where $f=\frac{e^2}{4\pi\epsilon_0\epsilon_\text{eff}}$ is a prefactor,
$\epsilon_\text{eff}$ is the effective dielectric constant of the hexagonal
boron nitride substrate, $r$ is the planar separation of $i$,$j$-th dipoles,
and $d$ is the interlayer distance. The interaction becomes a symmetric average
$V_{\text{Q}}(r) = [V_{\sigma\sigma} + V_{\sigma,-\sigma}]/2$ for the lower
quadrupolar exciton branch. The Hamiltonian is
\begin{equation}
  \mathcal{H} = \sum_{i < j} V(\mathbf{r}_{ij}) + \Big(\frac{\mathcal{E}^{+}_{q}-\mathcal{E}^{-}_{q}}{2}\Big)\, N_d
              + V_\text{M} N_{\text{M}},
\end{equation}
where the first term sums electrostatic interactions, $N_d$ counts
excitons in the dipolar state, and $N_{\text{M}}$ counts excitons
not trapped in a moiré.

\begin{figure}
\centering
\includegraphics[width=1\linewidth]{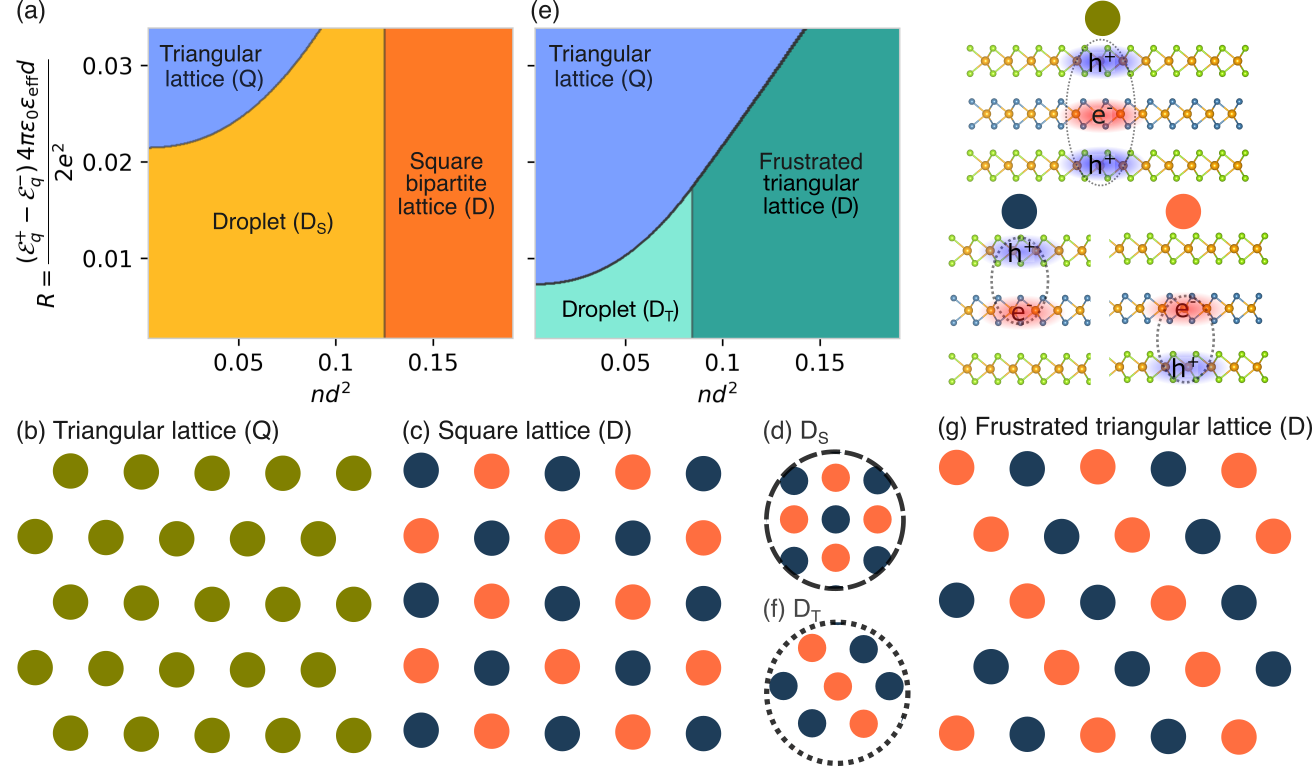}
\caption{\textbf{Many-exciton phase diagram for free and trapped quadrupolar
excitons.} (a)–(d): Free quadrupolar exciton phases include (b) a triangular
lattice of quadrupoles, (c) a square lattice of antiparallel dipoles, and (d)
droplet phases lacking long-range order. (e)-(g): Trapped quadrupolar exciton
phases include (e) a triangular lattice of quadrupoles, (g) frustrated
triangular lattice of anti-parallel dipoles, (f) droplet.} \label{fig5}
\end{figure}

Fig.~\ref{fig5}(a) shows the resulting phase diagram for free quadrupolar
excitons, computed with $\epsilon_{\text{eff}}=4$,
$\mathcal{E}_{+}-\mathcal{E}_{-}\leq20\,\text{meV}$, $d=0.61$ nm, and
$V_{\text{M}}=0$. We use the dimensionless parameters $nd^2,\text{and}\
R=\frac{(\mathcal{E}_{+}-\mathcal{E}_{-})\,4\pi\epsilon_0\epsilon_{\mathrm{eff}}d}{2e^2}$
to define the phase diagram, where $n$ is the exciton density. Three phases
emerge: a triangular lattice of quadrupolar excitons [Fig.~\ref{fig5}(b)], a
square
lattice of staggered dipolar excitons [Fig.~\ref{fig5}(c)], and an
inhomogeneous ``droplet" phase of dense dipolar clusters without long-range
order [Fig.~\ref{fig5}(d)]~\cite{Slobodkinquantum2020}. Quadrupolar excitons
repel via $V(r)\sim 1/r^5$, while antiparallel dipoles attract via $V(r)\sim
-1/r^3$. At low $n$, the electrostatic gain from conversion to staggered
dipoles is outweighed by $R$, stabilising the triangular quadrupolar lattice.
At high $n$, the gain exceeds $R$ and the system transitions to the staggered
dipolar phase, selecting a square bipartite geometry to avoid the frustration
inherent to the triangular lattice [Fig.~\ref{fig5}(e),(g)]. Moir\'{e} trapping
\textit{fundamentally} alters this competition by confining excitons to the
triangular lattice, expanding the quadrupolar phase boundaries
[Fig.~\ref{fig5}(e) with $V_{\text{M}}=30$ meV]. This resolves a
\textit{discrepancy} reported in Ref.~\citenum{Yuobservation2023}, where
quadrupolar excitons were observed at densities below $n \lesssim 3.3 \times
10^{13}\, \mathrm{cm}^{-2}$ ($nd^{2} \lesssim 0.12$). In this density range,
untrapped excitons with $R=0.02$ are expected to favour the dipolar droplet
phase (Fig.~\ref{fig5}(a)), making the observation of quadrupolar excitons
surprising. Moir\'{e} trapping stabilises the quadrupolar phase across this
range (Fig.~\ref{fig5}(e)), consistent with the experiment. While evidence for
staggered dipolar excitons has been found at higher
densities~\cite{Yuobservation2023}, the in-plane lattice ordering (for
instance, triangular versus square) has yet to be resolved experimentally. A
possible signature to distinguish them is the exchange-induced valley
splitting. Near $E_z = 0$, antiferroelectric correlations develop between the
dipolar excitons, suppressing the excitonic exchange interaction and reducing
the valley splitting~\cite{Liquadrupolar2023}. Since all nearest-neighbour
dipoles are antiparallel in the square phase but not in the frustrated
triangular phase, the magnitude of this suppression could distinguish the two
phases. A refined phase diagram including zero-point motion and
finite-temperature effects requires quantum Monte Carlo simulations or
finite-temperature unrestricted Hartree-Fock calculations incorporating the
moiré potential (see SI, Sec.~J for a discussion).

\section*{Discussion}
We develop a transfer-matrix formalism coupled with a fully
atomistic BSE framework, revealing the non-trivial real-space structure of
quadrupolar excitons in van der Waals trilayers. Moir\'{e} trapping gives rise
to two quadrupolar excitons with distinct internal structures, an azimuthally
symmetric state (I) and a threefold-symmetric state with a node at the hole
(II), which could serve as complementary probes of exciton correlations and of
the moir\'{e} potential that confines them. For example, the analogous
node-at-hole structure in H-stacked WS$_2$/WSe$_2$ heterobilayers stabilises a
long-lived, valley-polarised excitonic Mott state~\cite{Churobust2026}. Exciton
II lies $\sim$35~meV above I and couples to light more weakly, making it hard
to measure in low-temperature PL~\cite{Liquadrupolar2023}. It can be revealed
by electromodulation spectroscopy~\cite{Elyseoptical2022} or, potentially, by
PL at elevated temperature, where it becomes thermally populated. Moir\'{e}
trapping further establishes these excitons as a platform for exotic many-body
physics~\cite{Moessnergeometrical2006}, from frustrated quantum magnetism to
$SU(4)$ spin liquids.

\section*{Author contributions}
I.~M. and J.~L. discussed the initial project ideas. I.~M. led the project,
developed the algorithms, carried out the simulations, and wrote the first
draft of the manuscript. A.~A.~M. and J.~L. provided computing resources.
\'{A}.~R. contributed to discussions of the results. All authors contributed to
the manuscript.

\section*{Ethics declaration}
Not applicable.

\section*{Data availability}
All data used to produce the plots in the main text, and the atomic coordinates of
the optimized structures, are publicly available at
\url{https://github.com/imaitygit/quadrupolar-excitons} and at Zenodo
(\url{https://doi.org/10.5281/zenodo.24710175}).

\section*{Code availability}
Structure construction, atomic relaxations, electronic structure, and exciton
calculations presented in this work were performed using publicly available
codes. PyMEX is available at \url{https://github.com/imaitygit/PyMEX}. 

\section*{Acknowledgements}
This work used the ARCHER2 UK National Supercomputing Service
(https://www.archer2.ac.uk)~\cite{BeckettARCHER22024} through our membership of
the UKCP consortium and the HEC Materials Chemistry Consortium, both funded by
the Engineering and Physical Sicence Research Council [EP/X035891/1,
EP/X035859]. We thank Sufei Shi for comments on the
manuscript.

\section*{Methods} 

\noindent \textbf{Atomic structure.} Twisted bilayer WSe$_{2}$/WS$_{2}$
structures were generated using the \textsc{Twister}
package~\cite{Naiktwister2022}. In the trilayer system, the top and bottom
WSe$_{2}$ layers are aligned in parallel to form a symmetric moir\'{e}
superlattice.  Atomic relaxations were performed using classical interatomic
potentials in \textsc{Lammps}~\cite{lammps, Thompsonlammps2022}. Interactions
within each layer were modeled using the Stillinger–Weber
potential~\cite{Zhouhandbook2017}. Interactions between neighboring layers were
modeled using the Kolmogorov–Crespi potential~\cite{Naikkolmogorov2019}, as
these dominate over contributions from next-nearest-neighbor
layers~\cite{Maitytemeperature2018}.

\noindent \textbf{Electronic structure.} The \textsc{Siesta} package was
employed for electronic structure calculations using localized atomic orbitals
as basis functions~\cite{Solersiesta2002}. Norm-conserving Troullier--Martins
pseudopotentials~\cite{Troullierefficient1991} and the local density
approximation were used for the exchange-correlation
energy~\cite{Perdewself1981}. Wavefunctions were expanded in a double-$\zeta$
plus polarization basis. All moir\'{e} calculations used $\Gamma$-point
sampling, while lattice-matched system calculations employed a
$12\times12\times1$ $k$-point grid for the charge density, a plane-wave energy
cutoff of 100~Ry, an energy shift of 0.02~Ry, and a vacuum spacing of 20~\AA\
in the out-of-plane direction. Spin-orbit coupling~\cite{Fernandezseivane2006}
was excluded during Wannier function generation but was included to extract the
$\Gamma$--$K$ ordering of bands.

Wannier functions were generated via the one-shot projection
method~\cite{Marzarimaximally2012} as implemented in the \textsc{Wannier90}
code~\cite{Pizziwannier2020}, using $d$ and $p$ orbitals as initial
projections. Specifically, Kohn-Sham wavefunctions were projected onto
atom-centered $d$ ($d_{xy}$, $d_{yz}$, $d_{zx}$, $d_{x^2-y^2}$, $d_{z^2}$)
orbitals on every W atom and $p$ ($p_x$, $p_y$, $p_z$) orbitals on every S and
Se atom, followed by orthogonalization and
disentanglement~\cite{Souzamaximally2001}. We used minimal-distance replica
selection for Fourier interpolation of the Wannier
Hamiltonian~\cite{Pizziwannier2020, Maityorigin2026}. 

Spin-orbit coupling (SOC) was excluded from Wannier function generation, which
caused $\Gamma$-derived bands of the pristine unit cell to fold above
$K$-derived bands in the moiré Brillouin zone, incorrectly ordering the
$\Gamma_v \to K_c$ and $K_v \to K_c$ transition energies. We corrected this by
applying a rigid shift to $K$-derived valence bands identified at each
$\mathbf{k}$-point via the orbital-weight score \begin{equation}
\begin{aligned}
s_{n\mathbf{k}} = \sum_{\alpha} \Bigg(
& \sum_{m \in \{d_{x^2-y^2},\, d_{xy}\}}
\Big|C_{mn\mathbf{k}}^{\alpha}\Big|^2 \\
&-
\sum_{m \in \{d_{z^2},\, d_{xz},\, d_{yz}\}}
\Big|C_{mn\mathbf{k}}^{\alpha}\Big|^2
\Bigg),
\end{aligned}
\end{equation}
where $C_{mn\mathbf{k}}^{\alpha}$ are the Wannier projection coefficients for orbital $m$ on atom $\alpha$, summed over all atoms per layer. Bands with $s_{n\mathbf{k}} > 0.1$ were shifted by a rigid $\Delta E$ determined by explicit comparison to SOC-included calculations, restoring the correct $\Gamma$--$K$ ordering. This is physically motivated by the well-known orbital composition of 2D transition metal dichalcogenides valence bands: $K$ states are dominated by $d_{x^2-y^2}$ and $d_{xy}$, while $\Gamma$ states are dominated by $d_{z^2}$~\cite{Kormanyoskp2015}. We implemented this procedure in PyMEX. 

\noindent \textbf{Excitonic structure.} We implemented the transfer matrix
approach described above in \textsc{PyMEX}~\cite{Maityatomistic2025} for
arbitrarily stacked multi-layer structures. We employed \textsc{PyMEX} to
construct and diagonalise the BSE Hamiltonian and compute absorption spectra
using Wannier functions as basis, with diagonalisation accelerated by the
Eigenvalue soLvers for Petaflop Applications (\textsc{Elpa})
library~\cite{Auckenthalerparallel2011,Marekelpa2014}. We applied a vertical
electric field $E_{z}$ by adding the term $-eE_{z}\sum_{i} z_i |i\rangle
\langle i|$ to the electronic Hamiltonian, where $z_i$ is the $z$-coordinate of
the $i$-th Wannier function and $e$ is the electron charge. At each twist
angle, we performed BSE calculations at 25 electric-field values and
interpolated the results to construct a heatmap of the field-dependent
absorption spectra. We used $k$-point grids of $78\times78\times1$,
$18\times18\times1$, $6\times6\times1$, $6\times6\times1$, and
$3\times3\times1$, and included 3, 6, 20, 24, and 40 valence bands and 3, 6,
10, 8, and 16 conduction bands in the BSE Hamiltonian for the lattice-matched,
$16.1^\circ$, $4.7^\circ$, $2.8^\circ$, and $2.2^\circ$
WSe$_{2}$/WS$_{2}$/WSe$_{2}$ trilayers, respectively (See SI Sec.~H for the
convergence of the low-energy excitons with the number of moir\'{e} bands and
the $\mathbf{k}$-grid used to construct the BSE Hamiltonian). Spin-orbit coupling 
is omitted from our BSE Hamiltonian but
can be included perturbatively~\cite{Maityatomistic2025}. It splits the
quadrupolar $1s$-like exciton into singlet and triplet states separated by
$\sim$40\,meV, with the singlet being optically
bright~\cite{Deilmannquadrupolar2024}. All results presented in the main text
apply to both branches independently (See Sec.~G of the SI for a discussion of
the impact of spin–orbit coupling on quadrupolar excitons, calculated using the
spinor BSE). All our BSE calculations used the scissor-shifted DFT-level
electronic structure as input. See SI, Sec.~H for comments on the oscillations
in the dipole moments found in Fig.~4.

\textbf{Exciton phase diagram}. Lattice sums were evaluated over all integer
pairs $(n_1, n_2)$ with $|n_i| \leq 65$, excluding the origin, on both square
and triangular lattices, with lattice constants 
set by the site density $n$ via $a^2 = 1/n$ and $\tfrac{\sqrt{3}}{2}a^2 = 1/n$ 
respectively. See SI, Sec.~E for additional details. The dipolar phases carry an additional energy cost $R$ per exciton 
(the quadrupole-to-dipole gap), and the dipolar square lattice incurs a further moiré penalty 
$V_\mathrm{M}$ per site for being incommensurate with the triangular moiré lattice. 
Below their respective optimal densities $n^*$, each dipolar phase undergoes a 
droplet construction: excitons phase-separate into dense droplets at $n^*$ 
surrounded by empty regions, with the interaction energy fixed to its minimum 
value. Since moiré potentials can exceed $100\,\mathrm{meV}$~\cite{Maityelectrons2023,Rossianomalous2024}, the value 
$V_\mathrm{M} = 30\,\mathrm{meV}$ used here is conservative, and our 
conclusions remain robust. 

\bibliography{mem}

\clearpage
\section*{Figure legends}

\clearpage
\onecolumngrid
\setcounter{section}{0}
\setcounter{subsection}{0}
\setcounter{figure}{0}
\setcounter{table}{0}
\setcounter{equation}{0}
\renewcommand{\thesection}{\Alph{section}}
\renewcommand{\thesubsection}{\arabic{subsection}}
\renewcommand{\thefigure}{S\arabic{figure}}     
\renewcommand{\thetable}{S\arabic{table}}       
\renewcommand{\theequation}{S\arabic{equation}} 
\renewcommand{\theHsection}{SI.\Alph{section}}
\renewcommand{\theHsubsection}{SI.\Alph{section}.\arabic{subsection}}
\renewcommand{\theHfigure}{S\arabic{figure}}
\renewcommand{\theHtable}{S\arabic{table}}
\renewcommand{\theHequation}{S\arabic{equation}}

\begin{center}
{\large\bfseries Supplementary Information to ``Moir\'e trapping of quadrupolar excitons in van der Waals trilayers''}
\end{center}
\vspace{1em}

\section{Methods}

\subsection{Atomic structure and Wannier functions}
\begin{table}[h!]
\centering
\begin{tabular}{c c c }
\hline
Lattice constant (nm) & Number of atoms & Number of Wannier functions \\
\hline\hline 
0.325 (0$^\circ$) & 9 & 33 \\
1.15 (16.1$^\circ$) & 111 & 407 \\
3.49 (4.7$^\circ$) & 1029 & 3773 \\
4.89 (2.8$^\circ$) & 2013 & 7381 \\
5.59 (2.2$^\circ$) & 2625 & 9625 \\
0.325 (60$^\circ$) & 9 & 33\\
\hline
\end{tabular}
\caption{Summary of lattice constants, number of atoms, and number of Wannier functions used in the electronic structure calculations for the BSE.}
\label{SI_table1}
\end{table}

\subsection{Multilayer screening using the transfer matrix method}

\begin{figure}[ht!]
    \centering
    \includegraphics[scale=0.65]{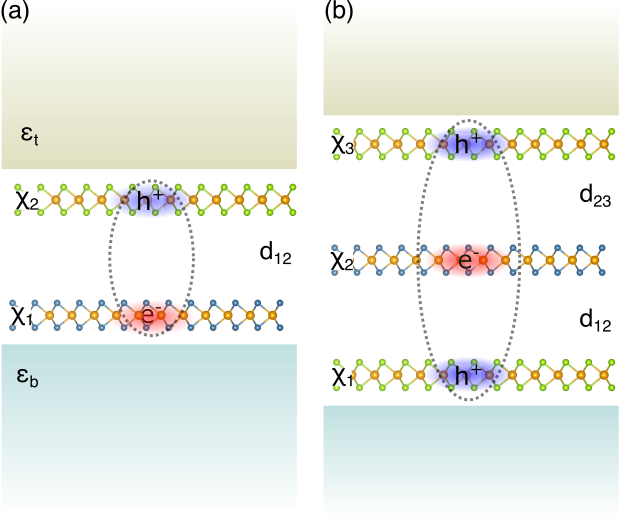}
    \caption{}
    \label{tm}
\end{figure}

\begin{itemize}
\item \textbf{Two stacked 2D materials}
\end{itemize}
Outside the stacked 2D materials, there are three regions: I ($z > d$), II ($0 < z < d$), and III ($z < 0$). We take $d = d_{12}$. The electrostatic potential admits the following solutions in each regions
\begin{equation}
\begin{split}
& \text{reg. I:}\  \phi_{I}(q, z) =  A e^{-qz} \\
& \text{reg. II:}\ \phi_{II}(q, z) = Be^{qz} + Ce^{-qz} \\
& \text{reg. III:}\ \phi_{III}(q, z) = De^{qz}
\end{split}
\label{waves_bilayer}
\end{equation}

We define a coefficient vector
$\psi_{z}= \begin{pmatrix}
\text{coeff. of}\ e^{-qz}\ \text{evaluated at}\ z \\ 
\text{coeff. of}\ e^{qz}\ \text{evaluated at}\ z \\ 
\end{pmatrix}$.  

\noindent \textbf{Propagation matrix:} Inside a homogeneous dielectric, the coefficients of the exponential basis evolve independently. Starting from
\begin{equation}
\psi(z)=
\begin{pmatrix}
U_{b}\\
V_{b}
\end{pmatrix},
\end{equation}
an out-of-plane shift by a distance $d$ gives
\begin{equation}
\psi(z+d)= \begin{pmatrix}
U_{t}\\
V_{t}
\end{pmatrix} =
\begin{pmatrix}
U_{b}e^{-qd}\\
V_{b}e^{qd}
\end{pmatrix}.
\end{equation}
This evolution can be captured compactly by a propagation matrix,
\begin{equation}
\psi(z+d)=P(d,\epsilon)\,\psi(z),
\end{equation}
where
\begin{equation}
\boxed{
P(d) =
\begin{pmatrix}
e^{-qd} & 0 \\
0 & e^{qd}
\end{pmatrix}
}
\label{PropMatrix}
\end{equation}

\noindent \textbf{Layer matrix:} The layer matrix is 2D polarizable, and the boundary conditions on $\phi(q, z)$ and $\frac{\partial \phi}{\partial z}$ need to be adjusted suitably. The setup consists of a 2D material with polarizability $\chi$, and dielectric constants $\epsilon_t$ and $\epsilon_b$ above and below it, respectively. Potenital continuity condition (i.e., $\phi(q,z\to 0^{-}) = \phi(q,z\to 0^{+})$) leads to
\begin{equation}
U_{b} + V_{b} = U_{t} + V_{t}
\label{continuity_layer}
\end{equation}
On the other hand, the discontinuous jump in the normal component of the electric field (i.e., $D_{z \to 0^{+}} - D_{z \to 0^{-}} = \sigma_{b} + \sigma_{f}$) leads to
\begin{equation}
-\epsilon_{0}\epsilon_{t}\frac{\partial \phi(q,z)}{\partial z}\Big|_{0^{+}} + \epsilon_{0}\epsilon_{b}\frac{\partial \phi(q,z)}{\partial z}\Big|_{0^{-}} = -\epsilon_{0}\chi q^{2}\phi(q, 0)
\end{equation}
In the above, we did not include source charge (i.e., $\sigma_{f}=0$). After a bit of algebra, this simplifies to 
\begin{equation}
\epsilon_{t}(U_{t} - V_{t}) - \epsilon_{b}(U_{b} - V_{b}) = -\chi q (U_{t} + V_{t})
\label{discontinuity_layer}
\end{equation}
Both Equations~\ref{continuity_layer} and~\ref{discontinuity_layer} must be satisfied simultaneously. Using the matrix formulation, both equations can be solved efficiently.
\begin{equation}
\begin{split}
& \begin{pmatrix}
1 & 1\\
\epsilon_{t} + \chi q  & -\epsilon_{t} + \chi q
\end{pmatrix} 
\begin{pmatrix}
U_{t} \\
V_{t}
\end{pmatrix} 
= \begin{pmatrix}
1 & 1 \\
\epsilon_{b} & - \epsilon_{b}
\end{pmatrix} 
\begin{pmatrix}
U_{b} \\
V_{b}
\end{pmatrix}  \\
&
\begin{pmatrix}
U_{t} \\
V_{t}
\end{pmatrix} 
= {\begin{pmatrix}
1 & 1\\
\epsilon_{t} + \chi q  & -\epsilon_{t} + \chi q
\end{pmatrix} }^{-1}
\begin{pmatrix}
1 & 1 \\
\epsilon_{b} & - \epsilon_{b}
\end{pmatrix} 
\begin{pmatrix}
U_{b} \\
V_{b}
\end{pmatrix} 
\end{split} 
\end{equation}
We can extract $\begin{pmatrix}
U_{t} \\
V_{t}
\end{pmatrix} = L(\chi, \epsilon_{b}, \epsilon_{t})\begin{pmatrix}
U_{b} \\
V_{b}
\end{pmatrix}$, where 
\begin{equation}
\boxed{
L(\chi, \epsilon_{b}, \epsilon_{t}) = \frac{1}{2\epsilon_{t}}\begin{pmatrix}
\epsilon_{t} + \epsilon_{b} -\chi q & \epsilon_{t} - \epsilon_{b} -\chi q \\
\epsilon_{t} - \epsilon_{b} +\chi q & \epsilon_{t} + \epsilon_{b} +\chi q
\end{pmatrix}
}
\label{Layer_without_source}
\end{equation}

\noindent \textbf{Source matrix:} Note that the source charge is not included in the above. The inclusion of source charge modifies Eqn.~\ref{discontinuity_layer}
\begin{equation}
\epsilon_{t}(U_{t} - V_{t}) - \epsilon_{b}(U_{b} - V_{b}) = -\frac{e}{\epsilon_{0}q}-\chi q (U_{t} + V_{t})
\label{discontinuity_with_source}
\end{equation}
This leads to different set of equations to solve
\begin{equation}
\begin{split}
& \begin{pmatrix}
1 & 1\\
\epsilon_{t} + \chi q  & -\epsilon_{t} + \chi q
\end{pmatrix} 
\begin{pmatrix}
U_{t} \\
V_{t}
\end{pmatrix} 
= \begin{pmatrix}
1 & 1 \\
\epsilon_{b} & - \epsilon_{b}
\end{pmatrix} 
\begin{pmatrix}
U_{b} \\
V_{b}
\end{pmatrix}  
+ 
\begin{pmatrix}
0 \\
-\frac{e}{\epsilon_{0}q}
\end{pmatrix} \\
&
\begin{pmatrix}
U_{t} \\
V_{t}
\end{pmatrix} 
= {\begin{pmatrix}
1 & 1\\
\epsilon_{t} + \chi q  & -\epsilon_{t} + \chi q
\end{pmatrix} }^{-1}
\begin{pmatrix}
1 & 1 \\
\epsilon_{b} & - \epsilon_{b}
\end{pmatrix} 
\begin{pmatrix}
U_{b} \\
V_{b}
\end{pmatrix} 
+ 
{\begin{pmatrix}
1 & 1\\
\epsilon_{t} + \chi q  & -\epsilon_{t} + \chi q
\end{pmatrix} }^{-1}
\begin{pmatrix}
0 \\
-\frac{e}{\epsilon_{0}q}
\end{pmatrix} \\
& 
\begin{pmatrix}
U_{t} \\
V_{t}
\end{pmatrix} =
\frac{1}{2\epsilon_{t}}\begin{pmatrix}
\epsilon_{t} + \epsilon_{b} -\chi q & \epsilon_{t} - \epsilon_{b} -\chi q \\
\epsilon_{t} - \epsilon_{b} +\chi q & \epsilon_{t} + \epsilon_{b} +\chi q
\end{pmatrix}
\begin{pmatrix}
U_{b} \\
V_{b}
\end{pmatrix} 
+ 
\begin{pmatrix}
-\frac{e}{2\epsilon_{0}\epsilon_{t}q} \\
\frac{e}{2\epsilon_{0}\epsilon_{t}q}
\end{pmatrix} \\
&
\begin{pmatrix}
U_{t} \\
V_{t}
\end{pmatrix} 
= L(\chi, \epsilon_{b}, \epsilon_{t})
\begin{pmatrix}
U_{b} \\
V_{b}
\end{pmatrix} 
+ 
S
\end{split} 
\end{equation}

The source matrix is given below, applies only at the source layer, and determines the propagated solutions thereafter.
\begin{equation}
\boxed{
S = \begin{pmatrix}
-\frac{e}{2\epsilon_{0}\epsilon_{t}q} \\
\frac{e}{2\epsilon_{0}\epsilon_{t}q}
\end{pmatrix} 
}
\end{equation}

We use the transfer matrix method to calculate the response of a multilayer system. Consider the bilayer 2D material setup shown in Fig.~\ref{tm}. In the bottom region ($z < 0$), there is a dielectric with constant $\epsilon_b$. Layer~1 is modeled as an ideal 2D layer at $z = 0$ with polarizability $\chi_1$. Between Layers 1 and 2 ($0 < z < d$), there is vacuum with $\epsilon = 1$. Layer~2 is an ideal 2D layer at $z = d$ with polarizability $\chi_2$. In the top region ($z > d$), a dielectric with constant $\epsilon_t$ is present. 

\textbf{Case I: Source charge in layer 1} 

The transfer matrix for the electrostatic potential, expressed in terms of the state vector $\psi_z$, can be written as
\begin{equation}
\begin{split}
\psi_{z\to \infty} & =  M \psi_{z\to -\infty}  \\
& = L(\chi_{2}, 1, \epsilon_{t})P(d,1)\big[L(\chi_{1}, \epsilon_{b}, 1)\psi_{z\to -\infty} + S\big] \\
\begin{pmatrix}
A \\
0
\end{pmatrix}& = \frac{1}{2\epsilon_{t}} 
\begin{pmatrix}
\epsilon_{t} + 1 -\chi_{2} q & \epsilon_{t} - 1 -\chi_{2} q \\
\epsilon_{t} - 1 +\chi_{2} q & \epsilon_{t} + 1 +\chi_{2} q
\end{pmatrix}
\begin{pmatrix}
e^{-qd} & 0 \\
0 & e^{qd}
\end{pmatrix}
\frac{1}{2}
\begin{pmatrix}
1 + \epsilon_{b} -\chi_{1} q & 1 - \epsilon_{b} -\chi_{1} q \\
1 - \epsilon_{b} +\chi_{1} q & 1 + \epsilon_{b} +\chi_{1} q
\end{pmatrix}
\begin{pmatrix}
0 \\
D 
\end{pmatrix} \\
& + 
\frac{1}{2\epsilon_{t}} 
\begin{pmatrix}
\epsilon_{t} + 1 -\chi_{2} q & \epsilon_{t} - 1 -\chi_{2} q \\
\epsilon_{t} - 1 +\chi_{2} q & \epsilon_{t} + 1 +\chi_{2} q
\end{pmatrix}
\begin{pmatrix}
e^{-qd} & 0 \\
0 & e^{qd}
\end{pmatrix}
\frac{e}{2\epsilon_{0}q}
\begin{pmatrix}
-1 \\
1
\end{pmatrix} 
\end{split}
\end{equation}

We obtain $D$ and propagate the solutions to extract $A,B,C$. 

\textbf{Case II: Source charge in layer 2} 
\begin{equation}
\begin{split}
\psi_{z\to \infty} & =  M \psi_{z\to -\infty}  \\
& = \big[L(\chi_{2}, 1, \epsilon_{t}) P(d,1)L(\chi_{1}, \epsilon_{b}, 1)\psi_{z\to -\infty} + S\big] \\
\begin{pmatrix}
    A \\
    0
\end{pmatrix}
& = \big[L(\chi_{2}, 1, \epsilon_{t}) P(d,1)L(\chi_{1}, \epsilon_{b}, 1) 
\begin{pmatrix}
    0 \\
    D
\end{pmatrix}
+ S\big]
\end{split}
\end{equation}

Again, $D$ is computed first, and the solutions are then propagated to determine the potentials in the other layer.

\begin{equation}
\boxed{\phi_{11} = \frac{C\left(\chi_2 q + \epsilon_t + 1 + \left(-\chi_2 q - \epsilon_t + 1\right)e^{-2d_{12}q}\right)}{q\left[\left(\chi_1 q + \epsilon_b - 1\right)\left(\chi_2 q + \epsilon_t - 1\right)e^{-2d_{12}q} - \left(\chi_1 q + \epsilon_b + 1\right)\left(\chi_2 q + \epsilon_t + 1\right)\right]}}
\end{equation}
\begin{equation}
\boxed{\phi_{12} = \frac{2C e^{-d_{12}q}}{q\left[
\left(\chi_1 q + \epsilon_b - 1\right)\left(\chi_2 q + \epsilon_t - 1\right)e^{-2d_{12}q}
- \left(\chi_1 q + \epsilon_b + 1\right)\left(\chi_2 q + \epsilon_t + 1\right)
\right]}}
\end{equation}
\begin{equation}
\boxed{\phi_{21} = \frac{2C e^{-d_{12}q}}{q\left[
\left(\chi_1 q + \epsilon_b - 1\right)\left(\chi_2 q + \epsilon_t - 1\right)e^{-2d_{12}q}
- \left(\chi_1 q + \epsilon_b + 1\right)\left(\chi_2 q + \epsilon_t + 1\right)
\right]}}
\end{equation}
\begin{equation}
\boxed{\phi_{22} = \frac{C\left[
\left(\chi_1 q + \epsilon_b + 1\right)
- \left(\chi_1 q + \epsilon_b - 1\right)e^{-2d_{12}q}
\right]}{q\left[
\left(\chi_1 q + \epsilon_b - 1\right)\left(\chi_2 q + \epsilon_t - 1\right)e^{-2d_{12}q}
- \left(\chi_1 q + \epsilon_b + 1\right)\left(\chi_2 q + \epsilon_t + 1\right)
\right]}}
\end{equation}

\vspace*{0.1 in}

\noindent \textbf{Sanity check and comparison with other works}: 
\begin{equation}
\begin{split}
\phi_{11} & = \frac{C\left[(r_2 q + 1) - r_2 q\, e^{-2d_{12}q}\right]}{2q\left[r_1 r_2 q^2 e^{-2d_{12}q} - (r_1 q + 1)(r_2 q + 1)\right]} \\
\phi_{21} & = \frac{C e^{-d_{12}q}}{2q\left[r_1 r_2 q^2 e^{-2d_{12}q} - (r_1 q + 1)(r_2 q + 1)\right]} \\
\phi_{21} & = \frac{C e^{-d_{12}q}}{2q\left[
r_1 r_2 q^2 e^{-2d_{12}q}
- \left(r_1 q + 1\right)\left(r_2 q + 1\right)
\right]} \\
\phi_{22} & =  \frac{C\left[\left(r_1 q + 1\right)- r_1 q\, e^{-2d_{12}q}\right]}{2q\left[r_1 r_2 q^2 e^{-2d_{12}q} - \left(r_1 q + 1\right)\left(r_2 q + 1\right)
\right]}
\end{split}
\end{equation}
In the limit where both sides are vacuum (i.e., $\epsilon_{t}=\epsilon_{b}=1$), our results are consistent with those of Danovich and co-workers~\cite{Danovichlocalized2018}. We have used $\chi_{i}=2r_{i}$. 

\begin{itemize}
\item \textbf{Three stacked 2D materials}
\end{itemize}

\noindent There are four regions of interest: region I ($z > d_{12} + d_{23}$), region II ($d_{12} < z < d_{12} + d_{23}$), 
region III ($0 < z < d_{12}$), and region IV ($z < 0$). Figure~\ref{tm} shows the setup. The electrostatic potential in these regions admits solutions of the form
\begin{equation}
\begin{split}
& \text{reg. I:}\  \phi_{I}(q, z)  =  A e^{-qz} \\
& \text{reg. II:}\ \phi_{II}(q, z)  = B^{qz} + Ce^{-qz} \\
& \text{reg. III:}\ \phi_{III}(q, z)  = De^{qz} + Ee^{-qz} \\
& \text{reg. IV:}\ \phi_{IV}(q, z)  = Fe^{qz}
\end{split}
\label{waves_trilayer}
\end{equation}
Three 2D materials have polarizabilities $\chi_1$, $\chi_2$, and $\chi_3$. Using the same approach as for the bilayer, we construct the electrostatic potential with the transfer matrix method. The source can be located in any of the three layers.

\noindent \textbf{Case I: Electron in layer 1 }
\begin{equation}
\begin{split}
\psi_{z\to \infty} & =  \mathcal{M} \psi_{z\to -\infty}  \\
& = L(\chi_{3}, 1, \epsilon_{t})P(d_{23},1)L(\chi_{2}, 1, 1)P(d_{12},1)\Big[L(\chi_{1}, \epsilon_{b}, 1)\psi_{z\to -\infty} + S\Big] \\
& = \begin{pmatrix}
M_{11} & M_{12} \\
M_{21} & M_{22}
\end{pmatrix}
\psi_{z\to -\infty} + \Big[ L(\chi_{3}, 1, \epsilon_{t})P(d_{23},1)L(\chi_{2}, 1, 1)P(d_{12},1) S\Big] \\
& = \begin{pmatrix}
M_{11} & M_{12} \\
M_{21} & M_{22}
\end{pmatrix}
\psi_{z\to -\infty} + \tilde{S} \\
\begin{pmatrix}
A \\
0
\end{pmatrix}& = \begin{pmatrix}
M_{11} & M_{12} \\
M_{21} & M_{22}
\end{pmatrix} 
\begin{pmatrix}
0 \\
F 
\end{pmatrix}
+
\begin{pmatrix}
\tilde{S}_{1} \\
\tilde{S}_{2} 
\end{pmatrix}
\end{split}
\end{equation}

We extract $F=-\frac{\tilde{S}_{2}}{M_{22}}$. As a sanity check, we verify that the left-hand side and right-hand side match after obtaining the solution. We propagate the solution, $F$ to obtain $\phi_{11},\ \phi_{12},\ \phi_{13}$. 
\begin{equation}
\begin{split}
\begin{pmatrix}
\phi_{11}^{1} \\
\phi_{11}^{2}
\end{pmatrix}
& = \Big[L(\chi_{1}, \epsilon_{b}, 1)\begin{pmatrix}
0 \\
F
\end{pmatrix}+ S\Big] \\
\begin{pmatrix}
\phi_{12}^{1} \\
\phi_{12}^{2}
\end{pmatrix}
& = L(\chi_{2}, 1, 1)P(d_{12},1) \Big[L(\chi_{1}, \epsilon_{b}, 1)\begin{pmatrix}
0 \\
F
\end{pmatrix}+ S\Big] \\
\begin{pmatrix}
\phi_{13}^{1} \\
\phi_{13}^{2}
\end{pmatrix}
& = L(\chi_{3}, 1, \epsilon_{t})P(d_{23},1) L(\chi_{2}, 1, 1)P(d_{12},1) \Big[L(\chi_{1}, \epsilon_{b}, 1)\begin{pmatrix}
0 \\
F
\end{pmatrix}+ S\Big] 
\end{split}
\end{equation}

\noindent \textbf{Case II: Electron in layer 2}
\begin{equation}
\begin{split}
\psi_{z\to \infty} & =  \mathcal{M} \psi_{z\to -\infty}  \\
& = L(\chi_{3}, 1, \epsilon_{t})P(d_{23},1)\Big[L(\chi_{2}, 1, 1) P(d_{12},1)L(\chi_{1}, \epsilon_{b}, 1)\psi_{z\to -\infty} + S\Big] \\
& = \begin{pmatrix}
M_{11} & M_{12} \\
M_{21} & M_{22}
\end{pmatrix}
\psi_{z\to -\infty} + \Big[ L(\chi_{3}, 1, \epsilon_{t})P(d_{23},1)S \Big] \\
& = \begin{pmatrix}
M_{11} & M_{12} \\
M_{21} & M_{22}
\end{pmatrix}
\psi_{z\to -\infty} + \tilde{S} \\
\begin{pmatrix}
A \\
0
\end{pmatrix}& = \begin{pmatrix}
M_{11} & M_{12} \\
M_{21} & M_{22}
\end{pmatrix} 
\begin{pmatrix}
0 \\
F 
\end{pmatrix}
+
\begin{pmatrix}
\tilde{S}_{1} \\
\tilde{S}_{2} 
\end{pmatrix}
\end{split}
\end{equation}

We extract $F=-\frac{\tilde{S}_{2}}{M_{22}}$. As a sanity check, we verify that the left-hand side and right-hand side match after obtaining the solution. Very similar to above, we propagate the solutions.

\noindent \textbf{Case III: Electron in layer 3}
\begin{equation}
\begin{split}
\psi_{z\to \infty} & =  \mathcal{M} \psi_{z\to -\infty}  \\
& = \Big[ L(\chi_{3}, 1, \epsilon_{t}) P(d_{23},1)L(\chi_{2}, 1, 1) P(d_{12},1)L(\chi_{1}, \epsilon_{b}, 1)\psi_{z\to -\infty} + S\Big] \\
& = \begin{pmatrix}
M_{11} & M_{12} \\
M_{21} & M_{22}
\end{pmatrix}
\psi_{z\to -\infty} + \mathbb{I}S \\
& = \begin{pmatrix}
M_{11} & M_{12} \\
M_{21} & M_{22}
\end{pmatrix}
\psi_{z\to -\infty} + \tilde{S} \\
\begin{pmatrix}
A \\
0
\end{pmatrix}& = \begin{pmatrix}
M_{11} & M_{12} \\
M_{21} & M_{22}
\end{pmatrix} 
\begin{pmatrix}
0 \\
F 
\end{pmatrix}
+
\begin{pmatrix}
\tilde{S}_{1} \\
\tilde{S}_{2} 
\end{pmatrix}
\end{split}
\end{equation}
We extract $F=-\frac{\tilde{S}_{2}}{M_{22}}$. Very similar to above, we propagate the solutions.

\noindent Explicit expressions are given below. Even with three layers the equations quickly become very complicated.
\begin{equation}
\boxed{
\begin{split}
\phi_{11} &= -\frac{C\left[\mathcal{A} + \mathcal{B}\,e^{-2d_{12}q}\right]}{\mathcal{N}} \\[6pt]
\phi_{12} &= \phi_{21} = \frac{2C\,\mathcal{C}\,e^{-d_{12}q}}{\mathcal{N}} \\[6pt]
\phi_{13} &= \phi_{31} = \frac{4C\,e^{-q(d_{12}+d_{23})}}{\mathcal{N}} \\[6pt]
\phi_{22} &= \frac{C\,\mathcal{C}\,\mathcal{D}}{\mathcal{N}} \\[6pt]
\phi_{23} &= \phi_{32} = \frac{2C\,\mathcal{D}\,e^{-d_{23}q}}{\mathcal{N}} \\[6pt]
\phi_{33} &= -\frac{C\left[\mathcal{E} + \mathcal{F}\,e^{-2d_{23}q}\right]}{\mathcal{N}}
\end{split}
}
\end{equation}
with 
\begin{equation}
\begin{split}
\mathcal{N} &= q\left[\mathcal{B}\left(\chi_1 q + \epsilon_b - 1\right)e^{-2d_{12}q}
               + \mathcal{A}\left(\chi_1 q + \epsilon_b + 1\right)\right] \\[6pt]
\mathcal{A} &= \chi_2 q\left(\chi_3 q + \epsilon_t - 1\right)e^{-2d_{23}q} 
               - \left(\chi_2 q + 2\right)\left(\chi_3 q + \epsilon_t + 1\right) \\[6pt]
\mathcal{B} &= \chi_2 q\left(\chi_3 q + \epsilon_t + 1\right) 
               - \left(\chi_2 q - 2\right)\left(\chi_3 q + \epsilon_t - 1\right)e^{-2d_{23}q} \\[6pt]
\mathcal{C} &= \left(\chi_3 q + \epsilon_t + 1\right) 
               - \left(\chi_3 q + \epsilon_t - 1\right)e^{-2d_{23}q} \\[6pt]
\mathcal{D} &= \left(\chi_1 q + \epsilon_b + 1\right) 
               - \left(\chi_1 q + \epsilon_b - 1\right)e^{-2d_{12}q} \\[6pt]
\mathcal{E} &= \chi_2 q\left(\chi_1 q + \epsilon_b - 1\right)e^{-2d_{12}q} 
               - \left(\chi_2 q + 2\right)\left(\chi_1 q + \epsilon_b + 1\right) \\[6pt]
\mathcal{F} &= \chi_2 q\left(\chi_1 q + \epsilon_b + 1\right) 
               - \left(\chi_2 q - 2\right)\left(\chi_1 q + \epsilon_b - 1\right)e^{-2d_{12}q}
\end{split}
\end{equation}

\noindent \textbf{Sanity checks:}

\noindent \textbf{Symmtery:} $\phi_{ij}=\phi_{ji}$. This is consistent with Green's reciprocity theorem. 

\noindent \textbf{Symmetric trilayer limit:} When $\chi_1=\chi_3$, $\epsilon_b=\epsilon_t$, $d_{12}=d_{23}$, we find
$\mathcal{A}=\mathcal{E},\quad\mathcal{B}=\mathcal{F},\quad\mathcal{C}=\mathcal{D} \text{ with }\phi_{11}=\phi_{33},\quad \phi_{12}=\phi_{23},\quad \phi_{22}\ \text{unchanged}$. 

\noindent \textbf{Exponetial decay of Off-diagonal elements :} $\phi_{12},\ \phi_{21} \sim e^{-d_{12}q}$, $\phi_{23},\ \phi_{32} \sim e^{-d_{23}q}$, and $\phi_{13},\ \phi_{31} \sim e^{-(d_{12}+d_{23})q}$.

\noindent \textbf{Bilayer limit:} i.e., ($d_{23}=0,\ \chi_3=0$) and examine the 
potentials to check consistency with bilayer calculations.
\begin{equation}
\phi_{11}\big|_{d_{23}=0,\,\chi_3=0} =
\frac{C\left[(\chi_2 q + \epsilon_t + 1) 
           - (\chi_2 q + \epsilon_t - 1)\,e^{-2d_{12}q}\right]}
{q\left[(\chi_2 q + \epsilon_t - 1)(\chi_1 q + \epsilon_b - 1)\,e^{-2d_{12}q}
      - (\chi_2 q + \epsilon_t + 1)(\chi_1 q + \epsilon_b + 1)\right]}
\end{equation}
\begin{equation}
\phi_{12}\big|_{d_{23}=0,\,\chi_3=0} =
\frac{2C\,e^{-d_{12}q}}
{q\left[(\chi_2 q + \epsilon_t - 1)(\chi_1 q + \epsilon_b - 1)\,e^{-2d_{12}q}
      - (\chi_2 q + \epsilon_t + 1)(\chi_1 q + \epsilon_b + 1)\right]}
\end{equation}
\begin{equation}
\phi_{13}\big|_{d_{23}=0,\,\chi_3=0} = \phi_{12}\big|_{d_{23}=0,\,\chi_3=0}
\end{equation}
\begin{equation}
\phi_{21}\big|_{d_{23}=0,\,\chi_3=0} = \phi_{12}\big|_{d_{23}=0,\,\chi_3=0}
\end{equation}
\begin{equation}
\phi_{22}\big|_{d_{23}=0,\,\chi_3=0} =
\frac{C\left[(\chi_1 q + \epsilon_b + 1) 
           - (\chi_1 q + \epsilon_b - 1)\,e^{-2d_{12}q}\right]}
{q\left[(\chi_2 q + \epsilon_t - 1)(\chi_1 q + \epsilon_b - 1)\,e^{-2d_{12}q}
      - (\chi_2 q + \epsilon_t + 1)(\chi_1 q + \epsilon_b + 1)\right]}
\end{equation}
\begin{equation}
\phi_{23}\big|_{d_{23}=0,\,\chi_3=0} = \phi_{22}\big|_{d_{23}=0,\,\chi_3=0}
\end{equation}
\begin{equation}
\phi_{31}\big|_{d_{23}=0,\,\chi_3=0} = \phi_{12}\big|_{d_{23}=0,\,\chi_3=0}
\end{equation}
\begin{equation}
\phi_{32}\big|_{d_{23}=0,\,\chi_3=0} = \phi_{22}\big|_{d_{23}=0,\,\chi_3=0}
\end{equation}
\begin{equation}
\phi_{33}\big|_{d_{23}=0,\,\chi_3=0} = \phi_{22}\big|_{d_{23}=0,\,\chi_3=0}
\end{equation}

These are consistent with the bilayer results.

\begin{itemize}
\item \textbf{n stacked 2D materials}
\end{itemize}

We follow the same strategy to solve the n-layer system. For example, when the source charge is in the bottom layer, we solve 
\begin{equation}
\begin{split}
\psi_{z\to \infty} & =  \mathcal{M} \psi_{z\to -\infty}  \\
& = L(\chi_{n}, 1, \epsilon_{t})P(d_{(n-1)n},1)\cdots\Big[L(\chi_{1}, \epsilon_{b}, 1)\psi_{z\to -\infty} + S\Big] \\
\begin{pmatrix}
A \\
0
\end{pmatrix}& = \begin{pmatrix}
M_{11} & M_{12} \\
M_{21} & M_{22}
\end{pmatrix} 
\begin{pmatrix}
0 \\
F 
\end{pmatrix}
+
\begin{pmatrix}
\tilde{S}_{1} \\
\tilde{S}_{2} 
\end{pmatrix}
\end{split}
\end{equation}

We extract $F=-\frac{\tilde{S}_{2}}{M_{22}}$. As a sanity check, we verify that the left-hand side and right-hand side match after obtaining the solution. We propagate the solution, $F$ to obtain $\phi_{11},\ \cdots \phi_{1n}$. And very similar to bilayer and trilayer system, we compute $\phi_{ij}$ where $i,j = 1,\ldots,n$.

\begin{itemize}
\item \textbf{Monolayer 2D material}
\end{itemize}
A special case of the transfer matrix method is the monolayer Keldysh potential. Using the same strategy as above, we set $n=1$. The potential becomes
\begin{equation}
   \phi_{11} =  -\frac{C}{q\left(2r_1 q + \epsilon_b + \epsilon_t\right)}
\end{equation}
This is the well-known Rytova-Keldysh potential in momentum space. 

\subsection{Exciton structure}

\clearpage
\newpage

\section{$K$-point convergence of dipolar and quadrupolar excitons}
\begin{figure}[ht!]
    \centering
    \includegraphics[width=\linewidth]{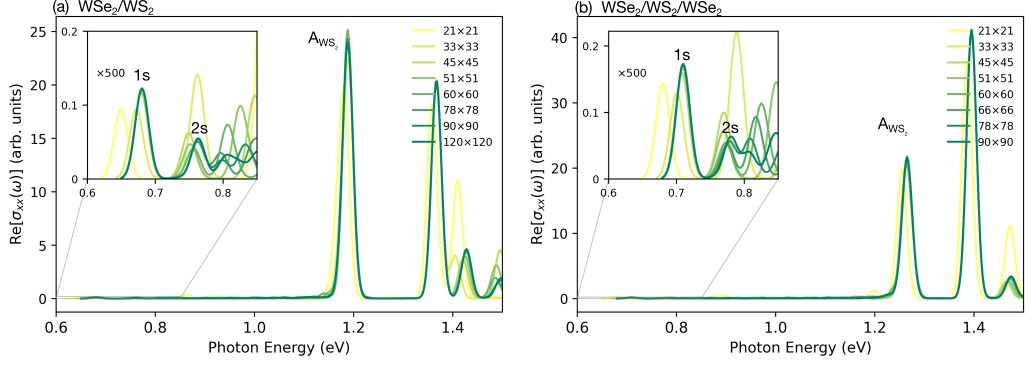}
    \caption{Convergence of the lowest-energy excitons in lattice-matched WSe$_2$/WS$_2$ and WSe$_2$/WS$_2$/WSe$_2$ van der Waals heterostructures with a lattice constant of 0.325~nm. We find that the 1s dipolar and quadrupolar excitons converge with a $45 \times 45$ $k$-point grid, while the 2s excitons require a $78 \times 78$ grid for convergence. We do not apply any rigid energy shifts. In constructing the BSE Hamiltonian, we include 2 valence and 2 conduction bands for the bilayer, and 3 valence and 3 conduction bands for the trilayer. We find that the oscillator strength of both dipolar and quadrupolar excitons is $\sim 10^{2}$ times weaker than that of intralayer excitons in WSe$_2$ or WS$_2$. The inset shows a plot zoomed in by a factor of 500.}
\end{figure}

\subsection{Convergence of lowest interlayer exciton energies}
Table~\ref{Convergence_1s} summarises the results for lattice matched system.
\begin{table}[ht]
\centering
\caption{Convergence of the lowest-lying $1s$ quadrupolar exciton in lattice matched system.}
\label{Convergence_1s}
\begin{tabular}{cc}
\hline
$\mathbf{k}$-grid & $1s$ quadrupolar exciton energy \\
\hline \hline 
 $21\times21\times1$ & 0.6801 eV \\
 $33\times33\times1$ & 0.7008 eV \\
 $45\times45\times1$ & 0.7085 eV \\ 
 $51\times 51\times1$ & 0.7095 eV \\
 $60\times60\times1$ & 0.7099 eV \\
 $66\times 66\times 1$ & 0.7100 eV \\
 $78\times78\times1$ & 0.7100 eV \\
 $90\times 90\times 1$ & 0.7100 eV \\ 
\hline
\end{tabular}
\end{table}

\begin{figure*}[h!]
\centering
\includegraphics[width=0.76\textwidth]{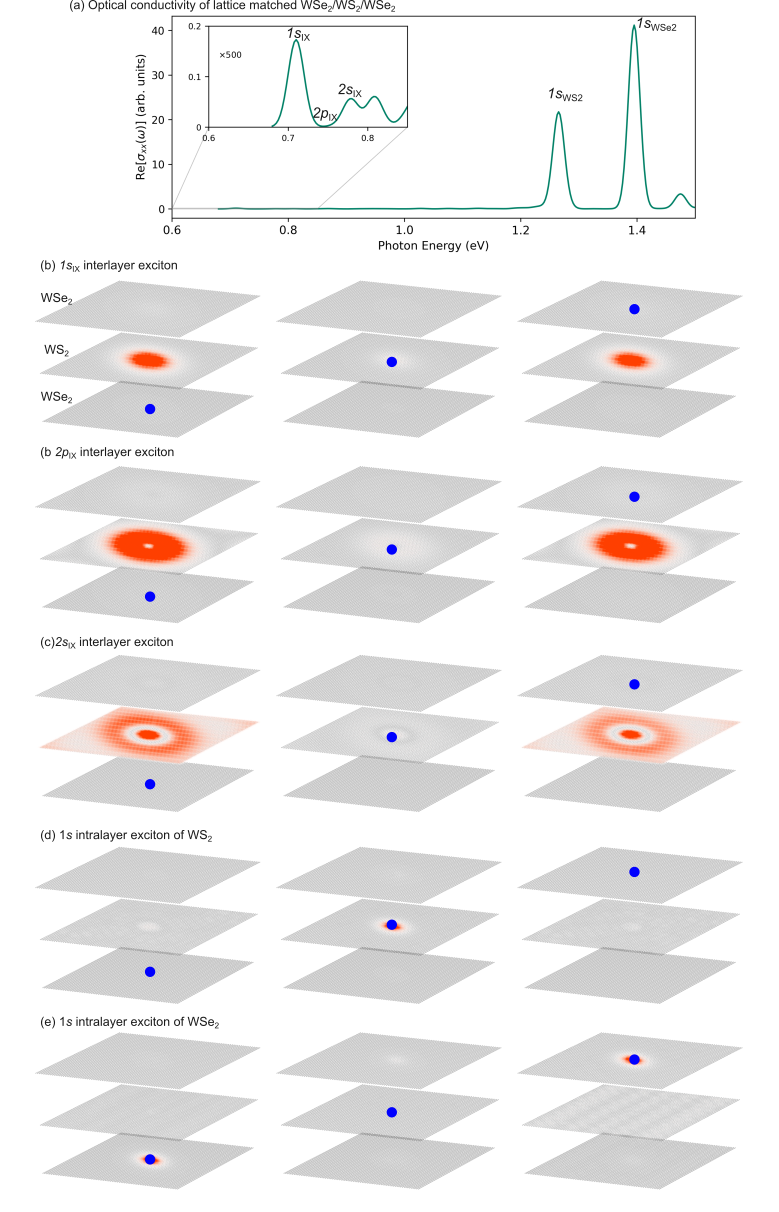}
\caption{\textbf{Exciton wavefunctions in lattice matched WSe$_2$/WS$_2$/WSe$_2$.} Blue dots indicate the hole positions. No energy shift was applied. Calculations neglected spin-orbit coupling, and the BSE Hamiltonian included 3 valence and 3 conduction bands on a $78\times78\times1$ $\mathbf{k}$-grid.}  
\label{exwfns_latticematched_R2}
\end{figure*}

\clearpage
\newpage

\section{Electrostatic potential for excitons I and II}
Exciton~I is the collinear arrangement $+q/2,\,-q,\,+q/2$ along $z$. Exciton~II replaces the central $-q$ by three charges $-q/3$ on an equilateral triangle of side $b$ in the $z=0$ plane. For exciton~I, with $k=1/4\pi\varepsilon_0$ and cylindrical coordinates $(\rho,z)$, superposing the Coulomb potentials gives
\begin{equation}
V(\rho,z) = k\left[
\frac{q/2}{\sqrt{\rho^{2}+(z-2d)^{2}}}
-\frac{q}{\sqrt{\rho^{2}+(z-d)^{2}}}
+\frac{q/2}{\sqrt{\rho^{2}+z^{2}}}
\right].
\end{equation}

We characterize both distributions through their multipole moments. The Cartesian quadrupole tensor, with $a$ labeling the point charges, is defined as
\begin{equation}
Q_{ij} = \sum_a q_a \left(3 x_{a,i} x_{a,j} - r_a^2 \delta_{ij} \right),
\end{equation}
and is traceless by construction ($Q_{ii}=0$). Equivalently, the moments are organized by the spherical multipole expansion,
\begin{equation}
q_{lm} = \sum_a q_a\, r_a^{\,l}\, Y_{lm}^{*}(\theta_a,\phi_a),
\end{equation}
with $l$ the multipole order and $m$ the azimuthal index (the quadrupole is $l=2$, the octupole $l=3$), and $Y_{lm}(\theta,\phi)\propto P_l^{m}(\cos\theta)\,e^{im\phi}$, where $P_l^{m}$ are the associated Legendre polynomials.

\textbf{Exciton I.} Measuring positions from the central $-q$ at $z=d$ ($z_a'=z_a-d$), the lower moments vanish:
\begin{equation}
\sum_a q_a = \tfrac{q}{2} - q + \tfrac{q}{2} = 0,
\end{equation}
\begin{equation}
\sum_a q_a z_a' = \tfrac{q}{2}(+d) - q(0) + \tfrac{q}{2}(-d) = 0.
\end{equation}
The charges lie on the axis, so $Q_{zz}=\sum_a q_a(3z_a'^2-r_a'^2)=2\sum_a q_a z_a'^2 = 2qd^2$, and tracelessness yields
\begin{equation}
Q_{xx} = Q_{yy} = -q d^2, \qquad Q_{zz} = 2 q d^2,
\end{equation}
so $Q_{xx}-Q_{yy}=0$. Thus, $q_{22}\propto(Q_{xx}-2iQ_{xy}-Q_{yy})=0$, and $q_{21}\propto (Q_{xz}-iQ_{yz})=0$. Also, note that $q_{l,-m}=(-1)^{m}q^{*}_{l,m}$~\cite{Jacksonclassical1998}. 

\textbf{Exciton II.} Centering at the origin (the midpoint of the hole charges and the centroid of the triangle), the in-plane charges sit at circumradius $R=b/\sqrt{3}$,
\begin{equation}
\mathbf{r}_1' = (0,\,R,\,0),\quad
\mathbf{r}_2' = \left(-\tfrac{\sqrt3}{2}R,\,-\tfrac{1}{2}R,\,0\right),\quad
\mathbf{r}_3' = \left(\tfrac{\sqrt3}{2}R,\,-\tfrac{1}{2}R,\,0\right).
\end{equation}
The monopole and dipole again vanish:
\begin{equation}
\sum_a q_a = \tfrac{q}{2} + \tfrac{q}{2} + 3\!\left(-\tfrac{q}{3}\right) = 0,
\end{equation}
\begin{equation}
\sum_a q_a\, \mathbf{r}_a' = \tfrac{q}{2}(0,0,d) + \tfrac{q}{2}(0,0,-d) - \tfrac{q}{3}\sum_i \mathbf{r}_i' = \mathbf{0}.
\end{equation}
The positive charges each contribute $\tfrac{q}{2}(2d^2)=qd^2$ to $Q_{zz}$, while each in-plane charge contributes $-\tfrac{q}{3}(-R^2)=\tfrac{q}{3}R^2$, so
\begin{equation}
Q_{zz} = 2qd^2 + qR^2 = 2qd^2 + \tfrac{1}{3}qb^2,
\end{equation}
giving $Q_{xx}=Q_{yy}=-qd^2-\tfrac{1}{6}qb^2$ and again $Q_{xx}-Q_{yy}=0$. As a result, $q_{l,m}$ remains unchanged. 

\subsection*{Octupole}
The two distributions first differ at the octupole ($l=3$). For exciton~I, all charges lie on the axis ($\theta_a=0,\pi$), where $Y_{3m}=0$ for $m\neq0$; the only possible component again vanishes,
\begin{equation}
q_{3,0}\propto\sum_a q_a z_a'^3 = \tfrac{q}{2}d^3 - q(0) + \tfrac{q}{2}(-d)^3 = 0,
\end{equation}
as enforced by the inversion center, which forbids all odd-$l$ moments. For exciton~II, the positive charges again give no octupole, while the triangle ($\theta_a=\tfrac{\pi}{2}$, $r_a=R$, $\phi_a=\phi_0+\tfrac{2\pi n}{3}$) gives
\begin{equation}
q_{3,m}\propto -\tfrac{q}{3}R^3\,P_3^{m}(0)\,e^{-im\phi_0}\sum_{n=0}^{2} e^{-i m\,2\pi n/3}.
\end{equation}
The azimuthal sum imposes a threefold selection rule, $\sum_{n=0}^{2} e^{-i m\,2\pi n/3} = 3\,\delta_{m\equiv 0\,(\mathrm{mod}\,3)}$, leaving $m\in\{0,\pm3\}$; since $P_3(0)=0$ and $P_3^3(0)=-15\neq0$, only
\begin{equation}
q_{3,\pm3}\propto q\,R^3 \propto q\,b^3 \neq 0
\end{equation}
remains. Exciton~II thus carries a threefold in-plane octupole, absent in exciton~I.

\clearpage
\newpage

\section{Structural relaxations in a 2.2$^\circ$ twisted WSe$_{2}$/WS$_{2}$/Wse$_{2}$ trilayer}
\begin{figure}[ht!]
    \centering
    \includegraphics[width=\linewidth]{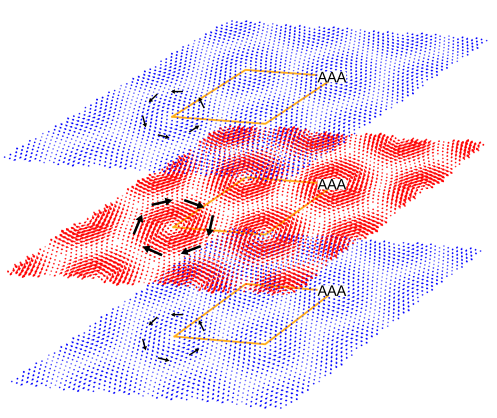}
    \caption{Large atomic relaxations in a 2.2$^\circ$ twisted WSe$_2$/WS$_2$/WSe$_2$ trilayer. Arrows show the displacement of W atoms in each layer relative to the rigid, unrelaxed structure. Displacements of W atoms in the middle WS$_2$ layer are amplified.}
\end{figure}

\clearpage
\newpage

\section{Electrostatic energies of dipolar and quadrupolar excitons on different lattice geometries}
\begin{figure}[ht!]
    \centering
    \includegraphics[width=0.75\linewidth]{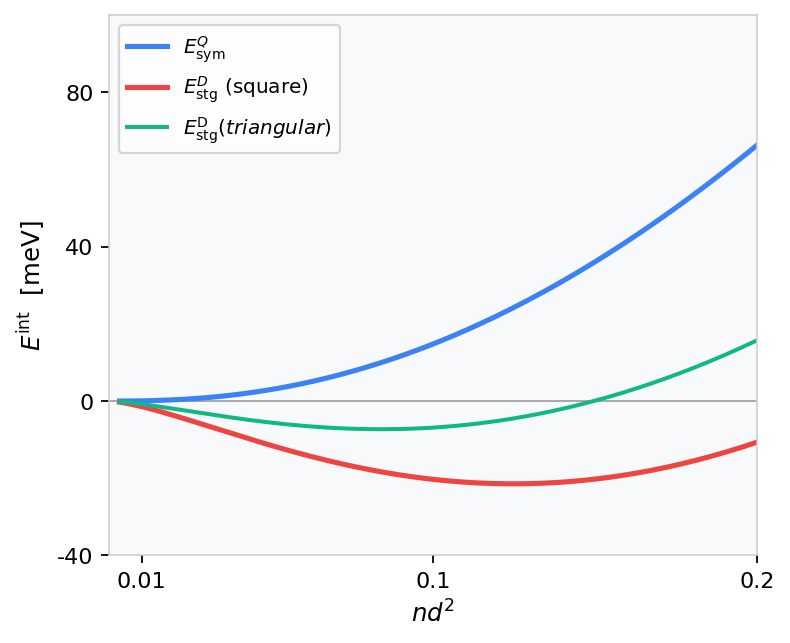}
    \caption{We consider three electrostatic energies: $E^{Q}_{\text{sym}}$, quadrupolar excitons on a triangular lattice; $E^{D}_{\text{stg}}$ (square), staggered dipolar excitons on a square lattice without frustration; and $E^{D}_{\text{stg}}$ (triangular), staggered dipolar excitons on a triangular lattice with minimal frustration.}
\end{figure}

\clearpage
\newpage

\section{In-plane strain development during relaxation and its impact on excitons}

\begin{figure*}[h!]
\centering
\includegraphics[width=1.0\textwidth]{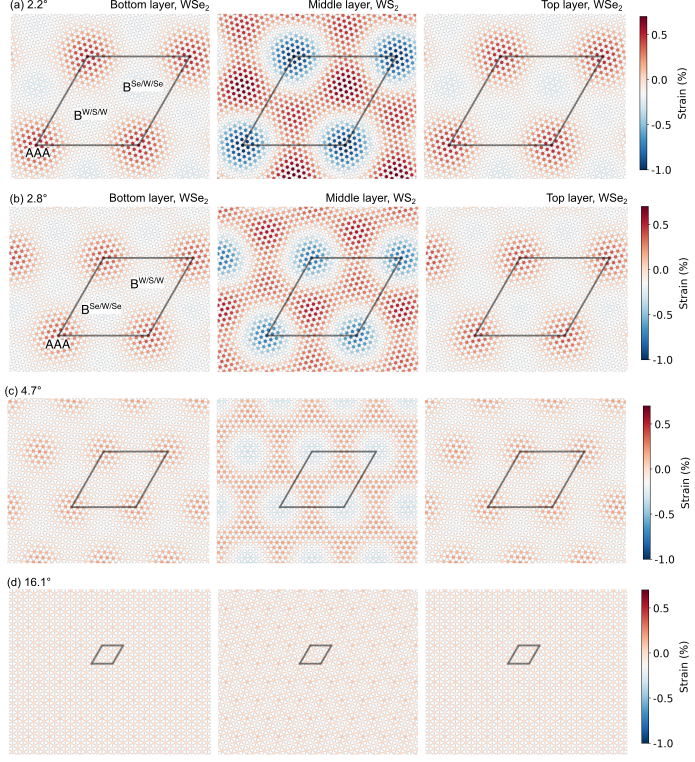}
\caption{Twist-angle dependent in-plane strain developed in each layer of WSe$_2$/WS$_2$/WSe$_2$ due to atomic rearrangements in the moir\'{e} pattern. The colorbar indicates the percentage of local strain, and the solid lines indicate the moir\'e  unit cell.}  
\label{strain2}
\end{figure*}

\begin{figure*}[h!]
\centering
\includegraphics[width=0.5\textwidth]{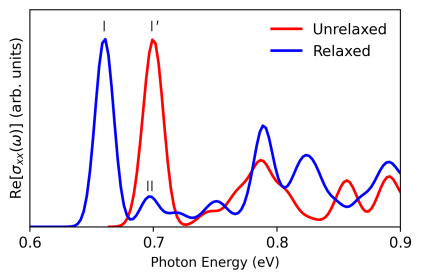}
\caption{Optical conductivities of relaxed and unrelaxed $2.2^\circ$ twisted WSe$_2$/WS$_2$/WSe$_2$ at zero external electric field. No rigid energy shift was applied.}  
\label{fig1_R1}
\end{figure*}

Fig.~\ref{fig1_R1} compares the optical conductivities of relaxed and unrelaxed WSe$_2$/WS$_2$/WSe$_2$ structures at a twist angle of $2.2^\circ$, with the outer WSe$_2$ layers aligned. We find that relaxation redshifts peak I (labeled I$^\prime$ in the unrelaxed structure) by about 38 meV and introduces a prominent peak II feature that is absent in the unrelaxed structure. 

\vspace*{0.2 cm}

\begin{figure*}[h!]
\centering
\includegraphics[width=\textwidth]{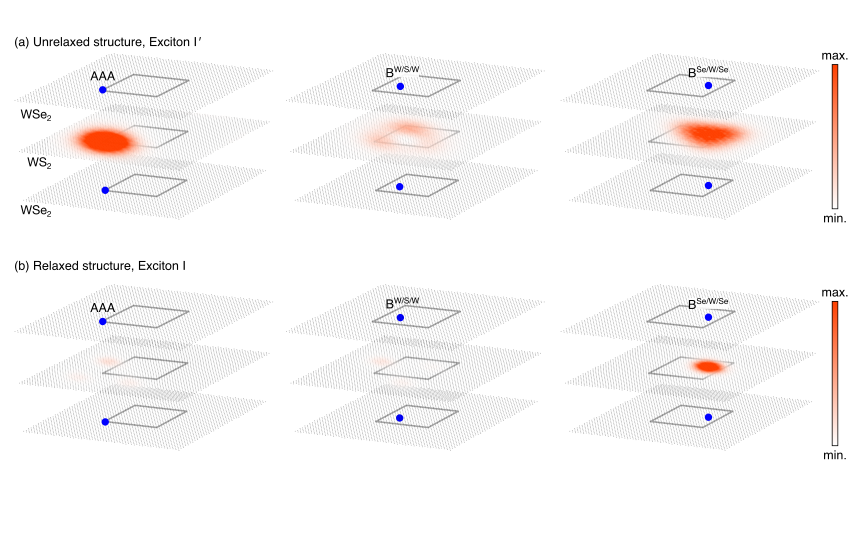}
\caption{Moiré quadrupolar exciton wavefunctions in a 2.2$^\circ$ twisted WSe$_2$/WS$_2$/WSe$_2$ trilayer for the (a) unrelaxed and (b) relaxed structures. Exciton I$^\prime$ does not show strong localization exclusively at the $\mathrm{B^{Se/W/Se}}$ region, unlike peak I in the relaxed structure. Blue dots indicate the hole positions, while solid lines show the moiré unit cell. The colorbar indicates the electron densities.}  
\label{fig2_R1}
\end{figure*}

Fig.~\ref{fig2_R1} shows the exciton wavefunction of the unrelaxed structure for peak I$^\prime$ at different hole positions, following the same analysis as for the relaxed structure presented in the main text. Importantly, we find that the exciton wavefunction in the unrelaxed structure shows electron density at both the AAA region and the $\mathrm{B^{Se/W/Se}}$ region. In contrast, peak I in the relaxed structure is strongly localized exclusively in the $\mathrm{B^{Se/W/Se}}$ region (Fig.~\ref{fig2_R1}(b)).

\subsection{Twist angle dependent lowest energy excitons}
\begin{figure*}[h!]
\centering
\includegraphics[width=\textwidth]{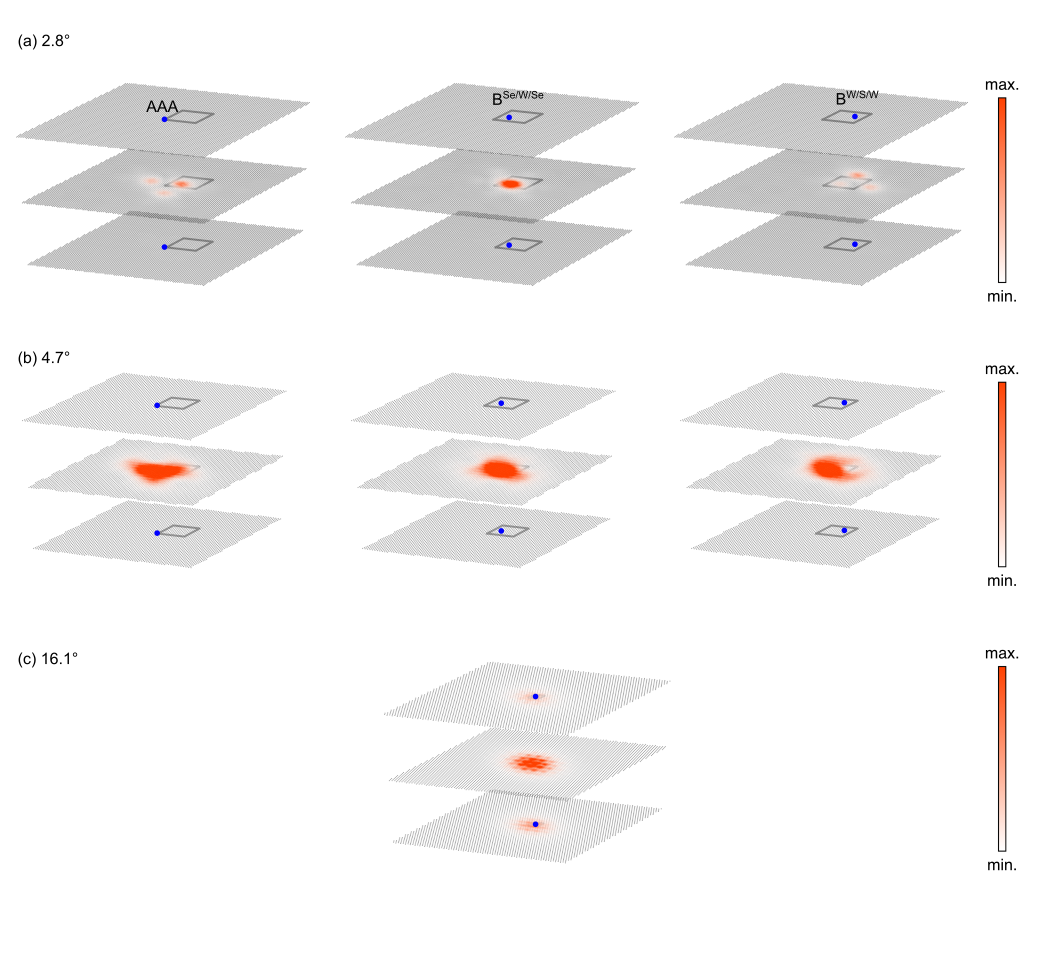}
\caption{\textbf{Moir\'{e} quadrupolar exciton wavefunctions in twisted WSe$_2$/WS$_2$/WSe$_2$.} Blue dots indicate the hole positions, while solid lines show the moir\'e unit cell. The colorbar indicates the electron densities. At $16.1^\circ$, the moiré unit cell is sufficiently small that distinguishing between different high-symmetry stackings is no longer meaningful. We therefore fix the hole at the origin.}  
\label{exwfns_R2}
\end{figure*}

Fig.~\ref{exwfns_R2} shows the lowest interlayer exciton wavefunctions, corresponding to the first peak in the optical conductivity. The exciton remains delocalised at large twist angles but becomes localised as the twist angle decreases. The 2.2$^\circ$ twisted trilayer is already included in the main text.

\clearpage
\newpage

\section{Impact of spin–orbit coupling on quadrupolar excitons}
\begin{figure*}[h!]
\centering
\includegraphics[width=\textwidth]{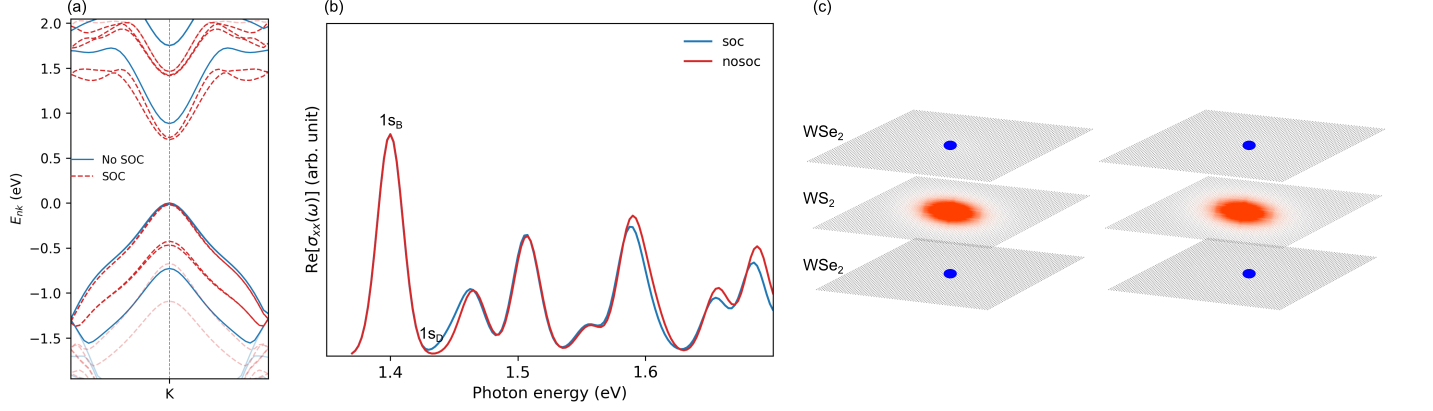}
\caption{\textbf{Electronic structure, optical conductivity, and exciton wavefunctions of lattice-matched WSe$_2$/WS$_2$/WSe$_2$ with spin–orbit coupling.} (a): Comparison of the electronic band structures near the high-symmetry K point, with and without spin–orbit coupling. The highlighted bands indicate those included in constructing the BSE Hamiltonian. (b): Comparison of the optical conductivities, with each spectrum normalised to unity. The lowest $1s$ exciton energy is set to 1.4 eV in each case. (c) Spin-split partners of the $1s$ exciton, with one optically bright ($1s_{\text{B}}$) and the other optically dark ($1s_{\text{D}}$). The blue dots indicate the hole locations in WSe$_2$.}  
\label{socvsnosoc}
\end{figure*}

We have performed extensive spinor Bethe–Salpeter equation calculations to quantify this effect and compared the resulting exciton wavefunctions with those obtained from spin-unpolarised calculations for lattice-matched WSe$_2$/WS$_2$/WSe$_2$. Fig.~\ref{socvsnosoc} summarise the results. Importantly, including spin–orbit coupling does not change the symmetry of the exciton wavefunctions. In the spin-unpolarised case, the lowest $1s$ quadrupolar exciton contains $n_v \times n_c \times n_{\mathbf{k}} = 2 \times 1 \times 2 = 4$ nearly degenerate BSE eigenstates. Here, $n_v = 2$ corresponds to the two valence bands of the WSe$_2$ layers, $n_c = 1$ to the conduction band of the WS$_2$ layer, and $n_{\mathbf{k}} = 2$ to the K and K$^\prime$ valleys. When spin–orbit coupling is included, the conduction band splits, with the two spin-split bands separated by only 23 meV. Consequently, the lowest $1s$ quadrupolar exciton contains approximately $2 \times 2 \times 2=8$ nearly degenerate BSE eigenstates. In contrast, the spin-split valence-band partner lies substantially lower in energy and does not contribute to the lowest-energy $1s$ quadrupolar exciton. 

\vspace*{0.2 cm}

We used Quantum ESPRESSO to compute the electronic structures with spin–orbit coupling and SIESTA for calculations without spin–orbit coupling.

\clearpage
\newpage

\section{Convergence with moir\'{e} bands and $\mathbf{k}$-grid}

There are two sets of parameters that control the convergence of lowest lying exciton energies: number of valence and conduction bands, and $\mathbf{k}$-grid used to construct the BSE Hamiltonian. We chose both parameters to converge the exciton states within $\sim$50~meV of the
lowest moir\'{e} exciton. 

\vspace*{0.2 cm}

\begin{figure*}[h!]
\centering
\includegraphics[width=0.9\textwidth]{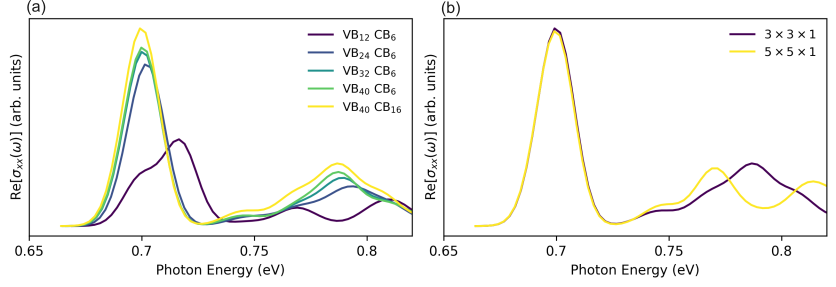}
\caption{\textbf{Convergence of the low-lying excitons in unrelaxed
2.2$^\circ$-twisted WSe$_2$/WS$_2$/WSe$_2$.} (a): Optical conductivity for different
numbers of moir\'{e} valence (VB) and conduction (CB) bands at a fixed
$3\times3\times1$ $\mathbf{k}$-grid. (b): Optical conductivity for $3\times3\times1$
and $5\times5\times1$ $\mathbf{k}$-grids with 40 valence and 16 conduction bands.}  
\label{convergence_bands_k}
\end{figure*}

(i) \textit{bands convergence:} As a test case, we vary the number of moir\'{e} valence and conduction bands at a fixed $3\times3\times1$ $\mathbf{k}$-grid and check convergence for unrelaxed 2.2$^\circ$-twisted WSe$_{2}$/WS$_{2}$/WSe$_{2}$. Fig.~\ref{convergence_bands_k}(a) summarizes our results. We find that the chosen parameters are sufficient to achieve convergence for the low-lying excitons of interest.  

\vspace*{0.2 cm}

(ii) \textit{$\mathbf{k}$-point convergence}: We provide a convergence analysis for the lattice-matched system in Sec.~B of the SI. We find that a $45\times45\times1$ $\mathbf{k}$-grid converges the energy of the lowest quadrupolar $1s$ exciton, whereas the more delocalised $2s$ state requires a denser $78\times78\times1$ grid because its larger real-space extent demands finer sampling in reciprocal space. In the main text, we focus on low-lying exciton states within 50~meV of the lowest-energy state of the moir\'{e} system. Table~\ref{convergence_k} summarises the $\mathbf{k}$-grids we use to construct the moir\'e BSE Hamiltonian, together with the corresponding effective unit-cell Brillouin-zone grids.

\begin{table}[ht]
\centering
\caption{Lattice constants and $\mathbf{k}$-point sampling used for the moiré Brillouin zone (BZ) and the equivalent unit-cell BZ.}
\begin{tabular}{cccc}
\hline
Twist angle & Lattice constant  & Moiré $\mathbf{k}$-grid & 
            Equivalent unit-cell $\mathbf{k}$-grid \\
\hline \hline 
16.1$^\circ$ & 1.15 nm & $18\times18\times1$ & $\sim 63\times63\times1$  \\
4.7$^\circ$ & 3.49 nm & $6\times6\times1$  & $\sim 64\times64\times1$ \\
2.8$^\circ$ & 4.89 nm & $6\times6\times1$ & $\sim 89\times89\times1$ \\
2.2$^\circ$ & 5.59 nm & $6\times6\times1$ & $\sim 51\times51\times1$ \\
\hline
\end{tabular}
\label{convergence_k}
\end{table}

As a test case, we vary the $\mathbf{k}$-grid with 40 valence and 16 conduction bands and check convergence for unrelaxed 2.2$^\circ$-twisted WSe$_{2}$/WS$_{2}$/WSe$_{2}$. Fig.~\ref{convergence_bands_k}(b) summarizes our results. We find that the chosen parameters are sufficient to achieve convergence for the low-lying excitons of interest. 

The oscillations in Fig.~4(a) are numerical rather than physical in origin. In moiré materials, we use a small fixed number of moiré electronic bands to construct the BSE Hamiltonian at each electric field. The number of bands was chosen to achieve reasonable convergence of the lowest BSE Hamiltonian energy while keeping the computational cost manageable, as we sample 25 different electric fields for each twist angle. As a result, even small convergence errors of $\sim 0.5$~meV are amplified when the dipole moment is obtained by numerical differentiation over finely spaced electric fields ($\Delta E_{z} = 4$~mV/nm). Specifically, an energy error of $\sim 0.5$~meV over this field interval corresponds to an uncertainty of $\sim 0.125$~e$\cdot$nm in the extracted dipole moment. This uncertainty appears as fluctuations in the computed dipole moments. 

\begin{figure*}[h!]
\centering
\includegraphics[width=\textwidth]{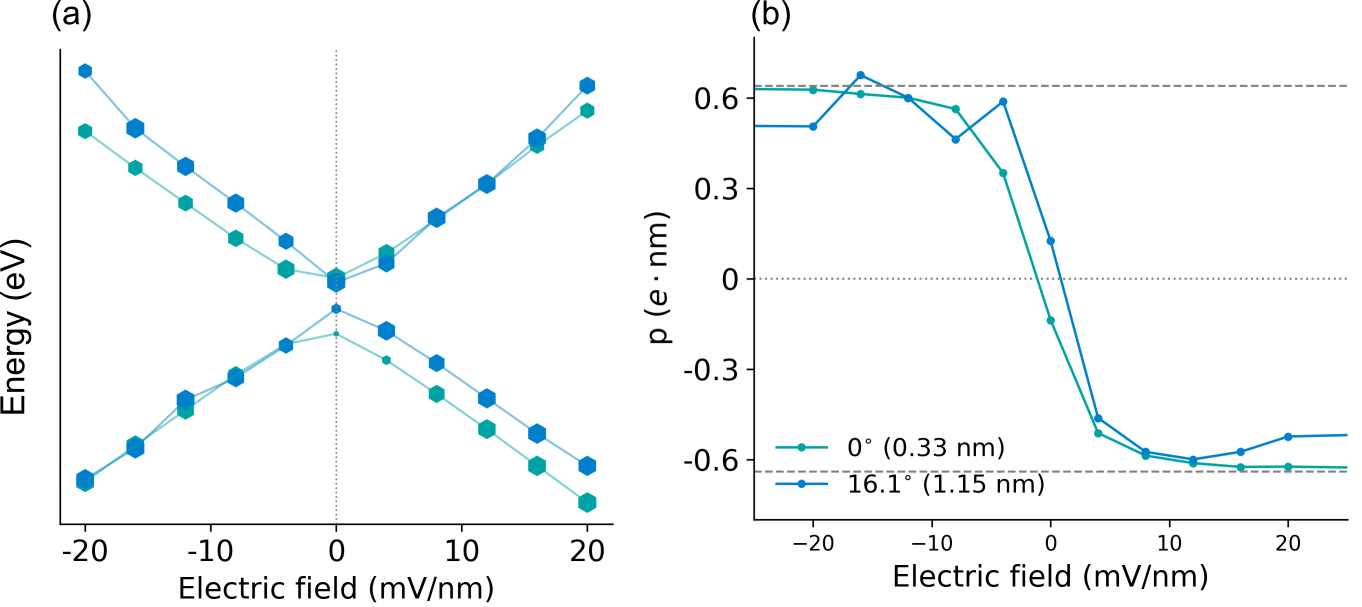}
\caption{\textbf{Avoided crossings and tunable dipole moments in twisted WSe$_{2}$/WS$_{2}$/WSe$_{2}$.} (a) $0^\circ$ lattice-matched trilayers have lower energy convergence errors, while $16.1^\circ$ moiré trilayers have higher convergence errors. This leads to larger fluctuations in their numerical derivatives, shown in (b) through the dipole moments.}  
\label{convergence}
\end{figure*}

Fig.~\ref{convergence} compares the $0^\circ$ lattice-matched and $16^\circ$ moiré structures of WSe$_2$/WS$_2$/WSe$_2$ (The same data are shown in Fig.~4 of the main text). The energies of the quadrupolar branches in the moiré structures appear reasonably smooth, but their numerical derivatives, and hence the dipole moments, show clear oscillations. In contrast, the unit-cell calculations do not show any pronounced oscillations. For the unit-cell calculation, we used 3 valence and 3 conduction bands to construct the BSE Hamiltonian, corresponding to valence and conduction bands within 1.3 eV of the valence band maximum and conduction band minimum, respectively. For the $16.1^\circ$ structure, we used 6 valence and 6 conduction bands to construct the BSE Hamiltonian, corresponding to valence and conduction bands within only 0.38 eV of the valence band maximum and 0.5 eV of the conduction band minimum, respectively. However, our main finding of an electric-field-driven quadrupolar-to-dipolar transition remains robust for all twist angles. Moreover, we already report the errors in the extracted quantities in Table~1 of the main text.

\clearpage
\newpage

\section{Fits of the avoided-crossing model to BSE dipole moments}
\begin{figure*}[h!]
\centering
\includegraphics[width=0.5\textwidth]{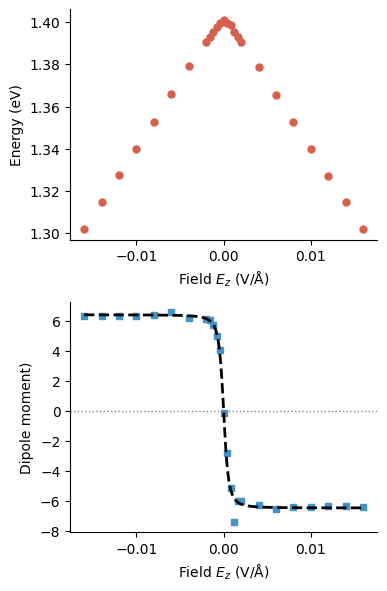}
\caption{Electric field dependence of the lowest quadrupolar exciton energies and their derivatives at 2.2$^\circ$ twisted WSe$_{2}$/WS$_{2}$/WSe$_{2}$. Data points are BSE calculations, and dashed lines denote the fits to the avoided-crossing model. Note that due to the fine $E_z$ steps near 0, small convergence errors in the exciton energies lead to oscillations (also discussed in a previous section).}  
\label{fit_quality}
\end{figure*}

Fig.~\ref{fit_quality} shows the electric-field dependence of the lowest quadrupolar exciton energies and their derivatives at 2.2$^\circ$ twisted WSe$_{2}$/WS$_{2}$/WSe$_{2}$ from our BSE calculations. We extract the dipole moment and tunnelling strength by fitting the model function, $p(E_z) = -\frac{d\mathcal{E}_{q}^{-}}{dE_z}
= \frac{p_0^{2} E_z}{\sqrt{(p_0 E_z)^2 + t^2}}$, to our BSE calculations. The model fits the data points well. The table in the main text already reports the uncertainties obtained from fitting the BSE results to the model. 

\clearpage
\newpage

\section{Comment on exciton phase diagram}
A detailed mapping of the phase diagram, including zero-point motion and finite-temperature effects, would require large-scale quantum Monte Carlo simulations incorporating the moir\'e potential. Unrestricted finite-temperature Hartree-Fock calculations offer a computationally cheaper, mean-field
alternative. Both lie beyond the scope of this work. Recent quantum Monte Carlo simulations by Zimmerman and co-workers include these effects~\cite{Zimmermanquadrupolar2026}. For experimentally
relevant parameters and in the absence of a moir\'e potential, they find that the quadrupole-to dipole change persists as a crossover, while the staggered square dipolar lattice is destabilised. Increasing the exciton mass, however, stabilises the ordered state: a staggered dipolar crystal emerges at larger exciton masses, \(m_X \gtrsim 4m_0\), and higher exciton densities. This
corresponds to a fourfold decrease in kinetic energy, which can be achieved at small twist angles, where moir\'e confinement flattens the exciton bands and thereby enhances the effective exciton
mass~\cite{Wutopological2017,Knorrexciton2022}. These results suggest that the ordered phases predicted with the simple model in the main text can be realised for strongly confined moir\'e excitons.

\clearpage
\newpage

\section{Exciton wavefunctions in 2.2$^\circ$ twisted WSe$_2$/WS$_2$/WSe$_2$.}

\begin{figure*}[h!]
\centering
\includegraphics[width=\textwidth]{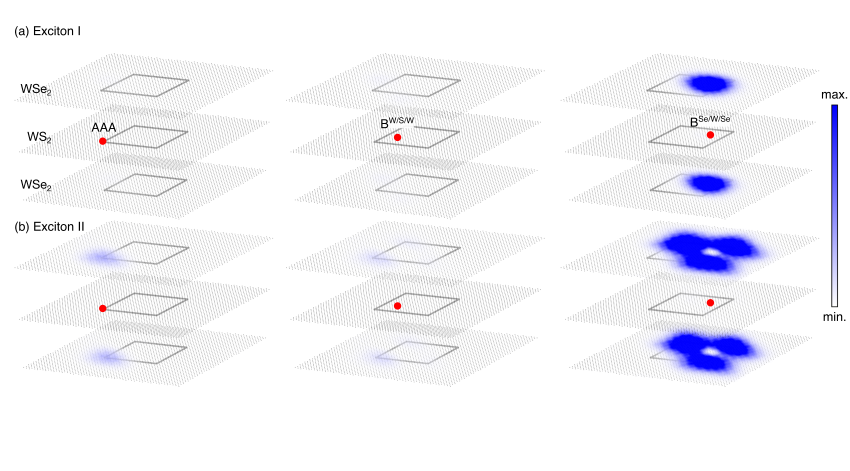}
\caption{\textbf{Exciton wavefunctions in 2.2$^\circ$ twisted WSe$_2$/WS$_2$/WSe$_2$.} Red dots indicate the electron positions. 
}  
\label{exwfns_fixedelectron_R2}
\end{figure*}

Fig.~\ref{exwfns_fixedelectron_R2} shows the hole distributions associated with Peak I and II excitons of 2.2$^\circ$ twisted WSe$_{2}$/WS$_{2}$/WSe$_{2}$. Associated electron distributions are in the main text, Fig.3. 

\clearpage
\newpage

\section{Twist-angle dependent tunable dipole moments and avoided crossings}

\begin{figure}[h!]
\centering
\includegraphics[width=1.0\linewidth]{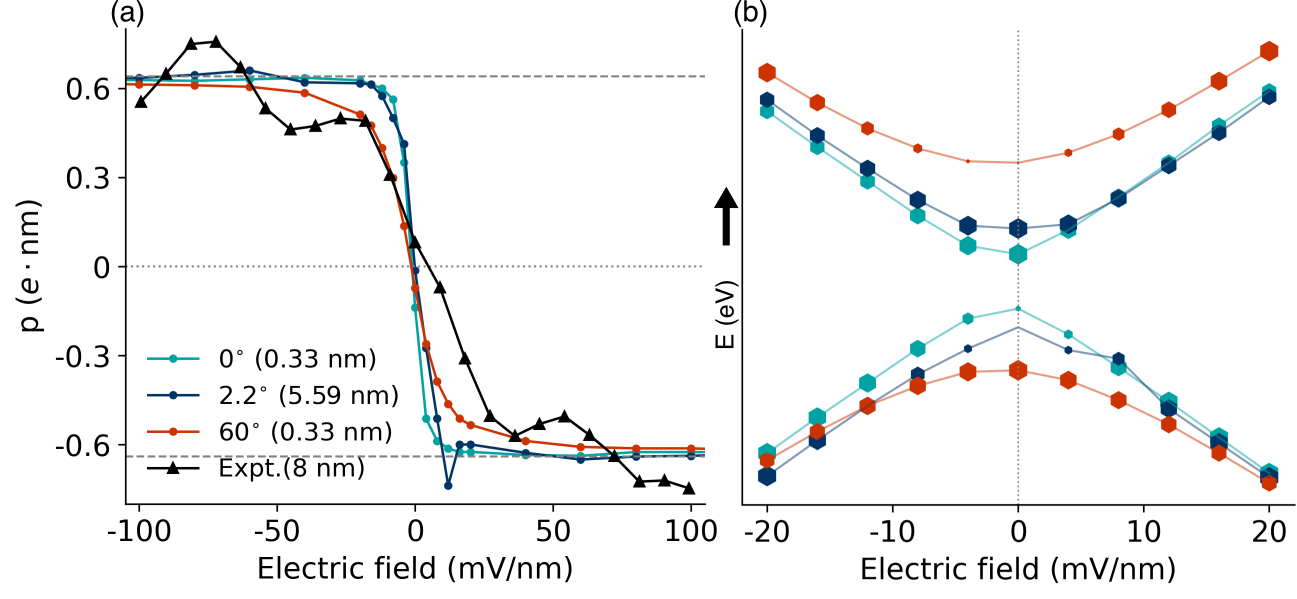}
\caption{\textbf{Tunable dipole moments and avoided crossings in twisted WSe$_{2}$/WS$_{2}$/WSe$_{2}$.} Field-dependent (a) dipole moments of the lower branch and (b) avoided-crossing gaps between the upper and lower branches for several twist angles. The center of the gap is aligned across all systems. Hexagon size indicates the oscillator strength of the excitons in (b). Experimental data adapted with permission from Yu \textit{et al}~\cite{Yuobservation2023} (Copyright 2023 Springer Nature).}
\label{SI_fig4}
\end{figure}

\end{document}